\documentclass[aps,pre,twocolumn,superscriptaddress,nofootinbib]{revtex4}

\usepackage{amsmath}
\usepackage{graphicx}
\usepackage{amssymb}
\usepackage{color}
\usepackage{mathrsfs}
\usepackage[colorlinks,linkcolor=blue,urlcolor=cyan,citecolor=blue,anchorcolor=blue]{hyperref}
\usepackage[T1]{fontenc}

\begin{document}	
\title{Different behaviors of diffusing diffusivity dynamics based on three
different definitions of fractional Brownian motion}

\author{Wei Wang}%\,\orcidlink{0000-0002-1786-3932}}
\affiliation{University of Potsdam, Institute of Physics \& Astronomy, 14476 
Potsdam-Golm, Germany}
\author{Qing Wei}%\,\orcidlink{0000-0001-5287-9430}}
\affiliation{LSEC, ICMSEC, Academy of Mathematics and Systems Science,
Chinese Academy of Sciences, Beijing 100190, China}
\author{Aleksei V. Chechkin}%\,\orcidlink{0000-0002-3803-1174}}
\affiliation{Institute of Physics \& Astronomy, University of Potsdam, 14476
Potsdam, Germany}
\affiliation{Faculty of Pure and Applied Mathematics, Hugo Steinhaus Center,
Wroc{\l}aw University of Science and Technology, 50-370 Wroc{\l}aw, Poland}
\affiliation{German-Ukrainian Core of Excellence, Max Planck Institute of
Microstructure Physics, Weinberg 2, 06120 Halle, Germany}
\affiliation{Asia Pacific Centre for Theoretical Physics, Pohang 37673, Republic
of Korea}
\author{Ralf Metzler}%\,\orcidlink{0000-0002-6013-7020}}
\email{rmetzler@uni-potsdam.de}
\affiliation{University of Potsdam, Institute of Physics \& Astronomy, 14476 
Potsdam-Golm, Germany}
\affiliation{Asia Pacific Centre for Theoretical Physics, Pohang 37673, Republic
of Korea}

\begin{abstract}
The effects of a "diffusing diffusivity" (DD), a stochastically time-varying
diffusion coefficient, are explored within the frameworks of three different
forms of fractional Brownian motion (FBM): (i) the Langevin equation driven
by fractional Gaussian noise (LE-FBM), (ii) the Weyl integral representation
introduced by Mandelbrot and van Ness (MN-FBM), and (iii) the Riemann-Liouville
fractional integral representation (RL-FBM) due to L{\'e}vy. The statistical
properties of the three FBM-generalized DD models are examined, including the
mean-squared displacement (MSD), mean-squared increment (MSI), autocovariance
function (ACVF) of increments, and the probability density function (PDF).
Despite the long-believed equivalence of MN-FBM and LE-FBM, their corresponding
FBM-DD models exhibit distinct behavior in terms of the MSD and MSI. In the
MN-FBM-DD model, the statistical characteristics directly reflect an effective
diffusivity equal to its mean value. In contrast, in LE-FBM-DD, correlations
in the random diffusivity give rise to an unexpected crossover behavior in both
MSD and MSI. We also find that the MSI and ACVF are nonstationary in RL-FBM-DD
but stationary in the other two DD models. All DD models display a crossover
from a short-time non-Gaussian PDF to a long-time Gaussian PDF. Our findings
offer guidance for experimentalists in selecting appropriate FBM-generalized
models to describe viscoelastic yet non-Gaussian dynamics in bio- and
soft-matter systems with heterogeneous environments.
\end{abstract}

\date{\today}

\maketitle

\section{Introduction}

Since Robert Brown first observed the erratic motion of micron-sized granules
ejected by pollen grains suspended in water \cite{brow1828}, this phenomenon,
now known as Brownian motion (BM), has been detected in numerous thermal
systems and has significantly influenced conceptual advances in nonequilibrium
statistical physics. A major breakthrough came with Albert Einstein's work
\cite{eins1905}, which provided a statistical interpretation of BM. Einstein
proposed that the particle position increments are independent random variables
following a specific distribution, leading to the prediction of the linear growth
of the mean-squared displacement (MSD) over time and a Gaussian probability
density function (PDF) for the displacement. Building on similar contributions
by William Sutherland \cite{suth1905}, Marian Smoluchowski \cite{smol1906},
and Paul Langevin \cite{lang1908}, the theory of Brownian motion was firmly
established. Notably, with the experimental work of Jean Perrin, sophisticated
experiments on Brownian motion allowed to pinpoint Avogadro's number and thus
to lay the foundation for the atomic theory of matter \cite{perrin}.

With recent advances of modern measurement techniques, in particular, single
particle tracking, anomalous diffusion with a power-law growth of the MSD \cite{metz2014}
\begin{eqnarray}
\label{msd}
\langle x^2(t)\rangle=2D t^{2H},
\end{eqnarray}
where $D$ is the generalized diffusion coefficient of physical dimension
$[D_{H}]=\mathrm{length}^2/\mathrm{time}^{2H}$ and $H$ is Hurst or anomalous
diffusion exponent, has been observed across a wide range of
spatiotemporal scales in various physical areas. Inter alia, these include
systems in cellular biology \cite{weis2004,gold2006,jeon2011,beta2024}, soft
matter \cite{jeon2016,gode2014,akim2011}, finance \cite{mich2003,mant2000},
ecology \cite{hump2010,meye2023,vilk2022a,vilk2022b}, astrophysics
\cite{yan2014}, geophysics \cite{brian,brian1}, and quantum physics
\cite{wei2022}. "Subdiffusion" features anomalous diffusion exponents in the
range $0<H<1/2$, while "superdiffusion" is realized for $1/2<H$. BM
corresponds to the special case $H=1/2$, and $H=1$ describes ballistic
transport \cite{bouchaud,igor,pt,metz2014,sikorski,mishura}.\footnote{The Hurst
exponent is used here due to its historical connotation with FBM. In anomalous
diffusion literature, another popular notation uses the anomalous diffusion
exponent $\alpha=2H$.}

Two prominent processes have proven particularly effective for the modeling of
anomalous diffusion across various systems. One is the continuous time random
walk \cite{mont1965,klablushle,hughes,report} with randomly distributed waiting
times $\tau$ between two successive jumps. When the associated waiting time PDF
$\psi(\tau)$ has the scale-free form $\psi(\tau)\simeq\tau^{-1-\alpha}$ with
$0<\alpha<1$ \cite{mandelbrot,harvey,klablushle,hughes,report}, the resulting
motion is subdiffusive with $H=\alpha/2$. The second common anomalous diffusion
process is fractional Brownian motion (FBM), initially introduced by Kolmogorov
\cite{kolm1940} and later formalized using stochastic integrals by Mandelbrot
and van Ness \cite{mand1968}. An alternative formulation for FBM is based on
the stochastic Langevin equation $dx(t)/dt=\sqrt{2D_{2H}}\xi_{H}(t)$, which is
driven by fractional Gaussian noise with the stationary autocovariance function
(ACVF) $\langle\xi_H(t)\xi_H(t+\tau)\rangle\sim2H(2H-1)D_H\tau^{2H-2}$ for
$\tau>0$ and $0<H\le1$. The ACVF is negative ("antipersistent") for subdiffusion
and positive ("persistent") for superdiffusion. FBM is widely used to model
diffusion in viscoelastic media \cite{weis2007,spak2010}, animal movement
patterns \cite{vilk2022b}, serotonergic brain fiber density profiles
\cite{janu2020,janu2023}, and roughness in financial data \cite{gath2018}.
FBM was studied under confinement by hard walls as well as potentials of the
generic form $V(x)\propto|x|^c$ ($c>0$), observing distinct non-equilibrium
shapes of the stationary PDF \cite{guggenberger,vojta,guggenberger1} as well
as multimodal states and non-confinement in shallow potential \cite{gugg,gugg1}.
More recently, FBM based on the fractional integral of Riemann-Liouville
type originally introduced by L{\'e}vy \cite{mand1968} with non-stationary
increments has garnered significant attention \cite{wang2023,balc2023,krap2024}
in systems with nonequilibrated initial condition, see also \cite{jakub}.

Diffusion in heterogeneous media displays numerous anomalous characteristics
\cite{waig2023}. In environments with "annealed" disorder,\footnote{Losely,
we distinguish "quenched" disorder, when the particle experiences the same
diffusivity value each time it revisits the same site, from annealed disorder,
when the value changes at each new visit \cite{bouchaud,agri,agri1,seongyu}.}
the particle's motion can be described by a purely time-dependent diffusion
coefficient, i.e., its diffusivity fluctuates over time. Such an annealed
heterogeneous dynamics leads to the phenomenon of Brownian yet non-Gaussian
motion, which has been observed in various complex systems \cite{wang2009,
wang2012,miot2021,alex2023,rusc2022,roic2020,seno2022}. In these processes,
while the MSD remains linear over time, the PDF deviates from the Gaussian
form, often exhibiting a distinct exponential shape that eventually crosses
over to a Gaussian distribution after a characteristic time (not all
experiments have a sufficiently large window to observe such a crossover).
The concept of diffusing diffusivity (DD), in which the diffusion coefficient
of the tracer particle itself becomes a time-dependent random process, was
first proposed by Chubinsky and Slater \cite{chub2014}. Chechkin et al.
\cite{chec2017} introduced a minimal DD model, in which the diffusion
coefficient evolves according to the square of an Ornstein-Uhlenbeck process.
Non-equilibrium initial conditions and other forms of random diffusivity
processes were considered in \cite{vittoria,vittoria1}. An alternative DD
formulation was presented by Tyagi and Cherayil \cite{tyagi}. Concurrently,
Jan and Sebastian \cite{jain2016,jain2017} formalize the DD model using a
path integral approach.

Viscoelastic and correlated superdiffusive yet non-Gaussian phenomena have
recently been observed in biological systems \cite{spak2017,tong2016,jeon2016,
beta2024,beta,amanda,thapa}, motivating the extension of the DD concept to
correlated processes such as FBM. In our earlier work \cite{wang2020a,wang2020b},
we introduced a
minimal FBM-generalized DD model formulated on the basis of the Langevin
equation description of FBM, exploring various diffusivity protocols. In
particular, unexpected crossovers in the MSD were observed beyond the correlation
time. More recently, FBM based on L{\'e}vy's Riemann-Liouville formulation with
a random diffusivity was analyzed in \cite{krap2024}---in this work no crossover
behavior in the MSD was found. Despite the similarities shared by different FBM
formulations, the combination with DD-driven dynamics turns out to effect significant differences. In this paper, we investigate FBM-generalized models
based on the three representations of FBM and analyze the impact of the random
diffusivity on the statistical dynamics, as quantified by the MSD, MSI, ACVF,
and the PDF.

The paper is organized as follows. In Sec. II, we introduce the physical
observables usually evaluated from experimental data. These will then be
employed to characterize the different DD models. In Sec. III, we present
the three definitions of FBM in detail and generalize them to FBM-DD models.
In Sec. IV, we present the numerical approach to generate trajectories of
the FBM-DD processes. Then we present the statistical characteristics of these
three FBM-DD models in Sec. V, including MSD, MSI, ACVF, and PDF. We summarize
and discuss our results in Sec. VI. We also present a table summarizing the
main results for comparison of the statistical properties of the three
FBM-generalized DD models.

\section{Physical observables}

To characterize the average diffusive behavior of tracer particles, the
conventional measurable is the MSD. It is calculated by averaging over an
ensemble of trajectories $x_i(t)$ at time $t$, relative to each particle's
initial position $x_i(0)$,
\begin{equation}
\langle x^2(t)\rangle=\frac{1}{N}\sum_{i=1}^N(x_i(t)-x_i(0))^2,
\end{equation}
where $N$ denotes the total number of trajectories.

The MSI, qualifying the displacement-increments during the lag time $\Delta$
starting at the physical time $t$, is defined as the mean-squared of the
increment in the form \cite{wei2025}
\begin{equation}
\label{ST}
\langle x^2_{\Delta}(t)\rangle=\langle[x(t+\Delta)-x(t)]^2\rangle.
\end{equation}
For processes with stationary increments, the MSI equals the MSD $\langle
x^2(\Delta)\rangle$ \cite{wei2025}. The MSI is equivalent to the structure
function originally introduced by Kolmogorov and Yaglom in their studies on
locally homogeneous and isotropic turbulence \cite{kolm1941a,kolm1941b,
yagl1987,yagl1953}.

The correlation of increments along an ensemble of time traces $x(t)$ can be
probed in terms of the ACVF
\begin{equation}
\label{eq-acf}
C^\delta(t,\Delta)=\delta^{-2}\langle x^\delta(t+\Delta)x^\delta(t)\rangle,
\end{equation}
with the increment $x^\delta(t)=x(t+\delta)-x(t)$. This ACVF is useful to
analyze the nature of the anomalous diffusion in the unconfined space
\cite{metz2014,igor}.

\section{FBM-generalized DD model}

\subsection{Representations of FBM}

In this section we introduce the three alternative forms of FBM we consider
in the following. They all encode the identical power-law scaling (\ref{msd})
of the MSD in their original formulation with constant parameters in unconfined
space. We then proceed to generalize these FBM processes by incorporating the
DD dynamics of their diffusivity, modeled as the square of an Ornstein-Uhlenbeck
process.

\subsubsection{Mandelbrot and van-Ness definition of FBM}

The most widely used representation of FBM, introduced by Mandelbrot and van-Ness
(MN-FBM) \cite{mand1968} and often referred to as "FBM I" in literature
\cite{mari99}, is defined for the Hurst exponent $0<H<1$ in the form
\begin{eqnarray}
\nonumber
x_{\mathrm{MN}}(t)&=&\sqrt{2DV_H}\Bigg\{\int_0^t(t-s)^{H-1/2}dB(s)\\
&&\hspace*{-1.2cm}+\int_{-\infty}^0\left[(t-s)^{H-1/2}-(-s)^{H-1/2}\right]d
B(s)\Bigg\},
\label{fbm1}
\end{eqnarray}
where $B(t)$ denotes a standard Brownian motion and $V_H$ is a constant given
by \cite{balc2022,mand1968}
\begin{eqnarray}
\nonumber
V_{H}&=&\left[(2H)^{-1}+\int_0^{+\infty}\left((1+z)^{H-1/2}-z^{H-1/2}\right)^2
dz\right]^{-1}\\
&=&\frac{\Gamma(2H+1)\sin(\pi H)}{\Gamma(H+1/2)^2}.  
\end{eqnarray}

\subsubsection{Langevin equation formulation of FBM}

An alternative formulation of MN-FBM, especially widely used in physics
literature, is defined through the overdamped Langevin equation, a process
we refer to as LE-FBM \cite{mand1968,metz2018},
\begin{eqnarray}
\label{fbm2}
\frac{dx_{\mathrm{LE}}(t)}{dt}=\sqrt{2D}\xi_H(t),
\end{eqnarray}
for $0<H\le1$. Here the driving fractional Gaussian noise has zero mean and ACVF
\begin{eqnarray}
\nonumber
\langle\xi^2_H\rangle_\Delta&=&\langle\xi_H(t+\Delta)\xi_H(t)\rangle\\
&=&\frac{1}{2\delta^2}\left(|\Delta+\delta|^{2H}+|\Delta-\delta|^{2H}
-2\Delta^{2H}\right).
\label{eq-fbm-acvf}
\end{eqnarray}. 

MN-FBM and LE-FBM are equivalent in the sense that they exhibiting the same
MSD and stationary MSI
\begin{equation}
\label{fbm3}
\langle x^2(\Delta)\rangle_{\mathrm{LE,MN}}=\langle x^{2}_{\Delta}(t)\rangle
_{\mathrm{LE,MN}}=2D\Delta^{2H}
\end{equation}
as well as the same stationary ACVF
\begin{equation}
\label{acf-mn-fbm}
C_{\mathrm{LE,MN}}^\delta(\Delta)=2D\left<\xi^2_H\right>_\Delta.
\end{equation}
In particular, when $\Delta\gg\delta$, the ACVF has the power-law decay
\begin{equation}
C_{\mathrm{LE,MN}}^\delta(\Delta)\sim 2DH(2H-1)\Delta^{2H-2}.
\end{equation}

\subsubsection{Riemann-Liouville formulation of FBM}

Recently, growing attention has been paid to an alternative definition of FBM
introduced by L\'evy \cite{mand1968,levy}, which is based on the Riemann-Liouville
fractional integral (RL-FBM) and referred to as the "FBM II" \cite{mari99}. It is
expressed as
\begin{eqnarray}
x(t)=\int_0^t\sqrt{4DH}(t-s)^{H-1/2}dB(s),
\end{eqnarray}
for $H>0$

RL-FBM shares the same MSD with MN-FBM and LE-FBM,
\begin{eqnarray}
\langle x^2(\Delta)\rangle_{\mathrm{RL}}=2D\Delta^{2H}.
\end{eqnarray}
However, the MSI of RL-FBM is nonstationary \cite{lim2001,wei2025}
\begin{eqnarray}
\label{rl-fbm-msi}
\langle x_\Delta^2(t)\rangle_\mathrm{RL}=4DH\Delta^{2H}\Bigg\{I_H\left(
\frac{t}{\Delta}\right)+\frac{1}{2H}\Bigg\},
\end{eqnarray}
where the integral $I_H(z)$ is given by
\begin{eqnarray}
I_H(z)=\int_0^z\left[(1+s)^{H-1/2}-s^{H-1/2}\right]^2ds. 
\end{eqnarray}
At short times $t\ll\Delta$, the MSI (\ref{rl-fbm-msi}) is approximately
identical to the MSD,
\begin{eqnarray}
\langle x^2(\Delta)\rangle_{\mathrm{RL}}\sim 2D\Delta^{2H}.  
\end{eqnarray}
At long times, $t\gg\Delta$, the value of the integral (\ref{rl-fbm-msi}) is 
given by
\begin{eqnarray}
I_H\left(\frac{t}{\Delta}\right)+\frac{1}{2H}\approx\frac{1}{V_H},
\end{eqnarray}
and the MSI becomes approximately stationary in this long time limit,
\begin{eqnarray}
\langle x^2_{\Delta}(t)\rangle_\mathrm{RL}\sim\frac{2D\Gamma(H+1/2)^2}{\Gamma(2H)
\sin(\pi H)}\Delta^{2H}.
\end{eqnarray}
It is worthwhile noting that the MSI of RL-FBM exhibits the same long-time
scaling behavior $\simeq\Delta^{2H}$ as the MSD but differs in its prefactor. 

The ACVF of RL-FBM is also nonstationary with the exact expression
\cite{wang2023,wei2025}
\begin{eqnarray}
\nonumber
C^{\delta}_\mathrm{RL}(t,\Delta)=2DH(2H-1)\hspace*{3.6cm}\\
\nonumber
\times\Big\{\frac{3-2H}{2}\Delta^{2H-2}\int_0^{t/\Delta}q^{H-1/2}(1+q)^{H-5/2}dq\\
+\delta^{-1}\Delta^{2H-1}\int_{t/\Delta}^{(t+\delta)/\Delta}q^{H-1/2}(1+q)^{
H-3/2}dq\Big\}.
\end{eqnarray}
In particular for $t=0$ the ACVF reads \cite{wei2025}
\begin{equation}
C^{\delta}_\mathrm{RL}(t,\Delta)\sim\frac{4DH(2H-1)\delta^{H-1/2}}{2H+1}\Delta^{
H-3/2}
\end{equation}
in the limit $\Delta\gg\delta$, and for long times $t\to\infty$, it becomes
stationary with
\begin{equation}
C_\mathrm{RL}^\delta(t,\Delta)\sim  \frac{2DH(2H-1)\Gamma\left(H+1/2\right)^2}{
\Gamma(2H)\mathrm{sin}(\pi H)}\Delta^{2H-2}.  
\end{equation}

For all three models discussed here, when $H=1/2$, the processes reduce to
standard BM. As indicated above, the three definitions are valid for different
ranges of the Hurst exponent $H$. Namely, for MN-FBM, the Hurst exponent needs
to be given by $0<H<1$, ensuring a positive prefactor $V_H>0$. For LE-FBM, the
Hurst exponent satisfies $0<H\le1$. In the case of RL-FBM, there are no strict
constraints on $H>0$. In this study, we specifically focus on the range $0<H<1$.

\subsection{Diffusing diffusivity}

The random diffusivity $D(t)$ in the following is assumed to follow the square
of an Ornstein-Uhlenbeck process $Y(t)$ \cite{chec2017},
\begin{subequations}
\begin{eqnarray}
\label{eq-diffusivity}
D(t)&=&Y^2(t),\\
\frac{d}{dt}Y(t)&=&-\frac{Y}{\tau}+\sigma\eta(t),
\end{eqnarray}   
\end{subequations} 
where $\eta(t)$ is a zero-mean white Gaussian noise with ACVF $\langle\eta(t)
\eta(t')\rangle=\delta(t-t')$, $D(t)$ and $Y(t)$ have physical units $[D]=
\mathrm{length^2}/\mathrm{time}^{2H}$ and $[Y]=\mathrm{length}/\mathrm{time}^H$.
Moreover, $\tau$ is the correlation or characteristic time of the
Ornstein-Uhlenbeck process, and $\sigma$ is the noise intensity for $D(t)$ with
units $[\sigma]=\mathrm{length}/\mathrm{time}^{H+1/2}$.

We assume equilibrium initial conditions for $Y(t)$, i.e., $Y(0)$ is taken
randomly from the equilibrium distribution
\begin{equation}
f_{\mathrm{eq}}(Y)=\frac{1}{\sqrt{\pi\sigma^2\tau}}\exp\left(-\frac{Y^2}{
\sigma^2\tau}\right).  
\end{equation}
Thus the process $Y(t)$ is stationary with the effective diffusivity
\cite{chec2017}
\begin{equation}
\langle D\rangle=\left\langle Y^2\right\rangle=\frac{\sigma^2\tau}{2}.    
\end{equation}

The second significant characteristic of the process $D(t)$ is the correlation
function of the square root  of $D(t)$ given by \cite{wang2020a}
\begin{eqnarray}
\nonumber
K(\Delta)&=&\left<\sqrt{D(t)D(t+\Delta)}\right>\\
\nonumber
&=&\frac{\sigma^2\tau}{\pi}\Big[\sqrt{1-e^{-2\Delta/\tau}}\\
&&\left.+e^{-\Delta/\tau}\arctan\left(\frac{e^{-\Delta/\tau}}{\sqrt{1-e^{-2\Delta/
\tau}}}\right)\right].
\label{eq-k_delta}
\end{eqnarray} 
When time is much shorter than the characteristic time, $\Delta\ll\tau$, the
diffusivity does not vary much and thus we have
\begin{equation}
\label{K1}
\langle D\rangle=\lim_{\Delta\to 0}K(\Delta)=\frac{\sigma^2\tau}{2}.   
\end{equation}
When time is much longer than the characteristic time $\Delta\gg \tau$, the 
correlations of the diffusivity decays exponentially. For this approximate
independence we then have
\begin{equation}
\label{K2}
K_{\mathrm{eff}}=\lim_{\Delta\to\infty}K(\Delta)=\langle|Y(t)|\rangle^2
=\frac{\sigma^2\tau}{\pi}. 
\end{equation}

\subsubsection*{Diffusing-diffusivity-generalized FBM}

We are now ready to define the three DD-generalized FBM models via introducing
the DD dynamics (\ref{eq-diffusivity}) based on the three FBM representations.\\

\paragraph*{\textbf{MN-FBM-DD:}} Combining the definition (\ref{fbm1}) of MN-FBM
with the DD dynamics (\ref{eq-diffusivity}), we obtain the MN-FBM-DD model:
\begin{eqnarray}
\label{model-b}
&&x_{\mathrm{MN}}(t)=\int_0^t\sqrt{2V_HD(s)}(t-s)^{H-1/2}dB(s)\nonumber\\
&&+\int_{-\infty}^0\sqrt{2V_HD(s)}\left[(t-s)^{H-1/2}-(-s)^{H-1/2}\right]d
B(s).\nonumber\\ 
\end{eqnarray}
In this case, the dynamics of the DD $D(t)$ is assumed to be at equilibrium at
all times $t$, including $t\le0$.

\paragraph*{\textbf{LE-FBM-DD:}} Adding the DD dynamics to the LE-FBM model
(\ref{fbm2}) leads us to the LE-FBM-DD model
\begin{eqnarray}
\label{model-a}
\frac{dx_{\mathrm{LE}}(t)}{dt}=\sqrt{2D(t)}\xi_H(t).   
\end{eqnarray}

\paragraph*{\textbf{RL-FBM-DD:}} Finally, from the definition (\ref{fbm3}) of
RL-FBM together with the DD dynamics we find the RL-FBM-DD process
\begin{eqnarray}
\label{model-c}
x_{\mathrm{RL}}(t)=\int_0^t\sqrt{4HD(s)}(t-s)^{H-1/2}dB(s).
\end{eqnarray}

\section{Simulations setup}

The discrete-time diffusivity $D(t_n)=Y^2(t_n)$ at time $t_n=n\times\delta t$,
where $\delta t$ is the time step, can be generated from the discretized
Ornstein-Uhlenbeck process (\ref{eq-diffusivity}),
\begin{equation}
Y(t_n)-Y(t_{n-1})=-\frac{1}{\tau}Y(t_{n-1})\delta t+\sigma\eta(t_{n-1})\delta t,
\end{equation}
where $\eta(t_n)=\eta_n/\sqrt{\delta t}$ and $\eta_n$ is a normally distributed
random variable with zero mean and unit variance.
 
\paragraph*{\textbf{LE-FBM-DD:}} For the LE-FBM-DD model, we discretize the
Langevin equation (\ref{model-a}) such that
\begin{eqnarray}
x_\mathrm{LE}(t_n)-x_\mathrm{LE}(t_{n-1})=\sqrt{2D(t_{n-1})}\xi_H(t_{n-1})
\delta t,
\end{eqnarray}
where $\xi_H(t_n)=\xi_{H,n}/(\delta t)^{1-H}$ and $\xi_{H,n}$ is the discrete
sequence of fractional Gaussian noise with zero mean and unit variance which
can be generated from standard approaches \cite{diek2004,hosk1984,inference}.
In this paper, we employ the Wood-Chan method \cite{wood1994} due to its rapid
simulation times achieved by using the discrete Fourier transformation.

\paragraph*{\textbf{RL-FBM-DD:}} For the RL-FBM-DD model, a direct approach
to discretize the stochastic integral may be adopted in the form
\begin{equation}
x_\mathrm{RL}(t_n)=\sum_{i=0}^{n-1}\sqrt{4HD(t_i)}\eta(t_i)\delta t\times
\left(t_n-t_i\right)^{H-1/2}.
\end{equation}
Typically, this direct approach is not very accurate for unveiling the scaling
of statistical quantities when $H<1/2$ as the weight $w(t_n-t_i)=(t_n-t_i)^{H-
1/2}$ becomes excessively large when $t_n-t_i$ is small.

Instead, we discretize the stochastic integral in the form
\begin{equation}
x_\mathrm{RL}(t_n)=\sum_{i=0}^{n-1}\int_{t_i}^{t_{i+1}}\sqrt{4HD(s)}\left(
t_n-s\right)^{H-1/2}\eta(s)ds.
\end{equation}
A simple approach can be applied when $D(s)\approx D(t_i)$ and $\eta(s)\approx
\eta(t_i)$ within the short time interval $[t_i,t_{i+1}]$. In this case we get
\begin{eqnarray}
\nonumber
x_\mathrm{RL}(t_n)&=&\sum_{i=0}^{n-1}\sqrt{4HD(t_i)}\eta(t_i)\int_{t_i}^{t_{i
+1}}\left(t_n-s\right)^{H-1/2}ds,\\
\nonumber
&=&\sum_{i=0}^{n-1}\sqrt{4HD(t_i)}\eta(t_i)\delta t\\
&&\times\frac{(t_n-t_i)^{H+1/2}-(t_n-t_{i+1})^{H+1/2}}{(H+1/2)\delta t}.
\end{eqnarray}
This form ensures that the power-law weight does not have any singularity.
Alternatively, the weight function
\begin{equation}
w(t_n-t_i)=\left[\frac{(t_n-t_i)^{2H}-(t_n-t_{i+1})^{2H}}{2H\delta t}\right]
^{1/2}.
\end{equation}
can be applied \cite{pina1994} to improve the prediction of the variance
scaling properties; in our simulations, we adopt this protocol.

\paragraph*{\textbf{MN-FBM-DD:}} Analogous to the RL-FBM-DD model, for the
MN-FBM-DD model the discrete strategy can be formulated as
\begin{eqnarray}
\nonumber
x_\mathrm{MN}(t_n)&=&\sum_{i=-n_a}^{n-1}\sqrt{2V_HD(t_i)}\eta(t_i)\delta t
\times w(t_n-t_i)\\
&&-\sum_{j=-n_a}^0\sqrt{2V_HD(t_j)}\eta(t_j)\delta t\,w(-t_j).
\end{eqnarray}
In practice, choosing $n_a=(T/\delta t)^{3/2}$ is sufficient \cite{diek2004,
inference}.

\section{Statistical properties of the DD-generalized FBM models}

\subsection{MSD}

\paragraph*{\textbf{LE-FBM-DD:}} The MSD of LE-FBM-DD model (\ref{model-a}) was
analyzed in \cite{wang2020a}, finding
\begin{equation}
\label{le-fbm-msd}
\langle x^2(t)\rangle_{\mathrm{LE}}=4\int_0^t(t-s)K(s)\langle\xi_H^2\rangle_sds.
\end{equation}
Here $K(s)$ and $\langle\xi_H^2\rangle_s$ are given by Eqs.~(\ref{eq-k_delta})
and (\ref{eq-fbm-acvf}). Expression (\ref{le-fbm-msd}) reveals an intriguing
crossover behavior: at short times $t\ll\tau$, the MSD reads
\begin{eqnarray}
\nonumber
\langle x^2(t)\rangle_\mathrm{LE}&\sim& 4K(0)\int_0^t(t-s)\langle\xi_H^2\rangle_s
ds\\
&\sim&2\langle D\rangle t^{2H}.
\label{le-fbm-dd-msd-st}
\end{eqnarray}
However, at long times $t\gg\tau$, different behaviors are emerging. For
superdiffusion ($H>1/2$), the MSD the scaling is identical to the short-time
scaling, but with a different prefactor,
\begin{equation}
\label{le-fbm-dd-msd-lt-1}
\langle x^2(t)\rangle_\mathrm{LE}\sim 2K_{\mathrm{eff}}t^{2H}.
\end{equation} 
In contrast, for subdiffusion ($H<1/2$), the MSD crosses over to normal
diffusion
\begin{equation}
\label{le-fbm-dd-msd-lt-2}
\langle x^2(t)\rangle_\mathrm{LE}\sim2D_{\mathrm{eff}}t.
\end{equation}
Here the effective diffusion coefficient $K_\mathrm{eff}$ is given by
Eq.~(\ref{K2}), and $D_\mathrm{eff}$ corresponds to the form
\begin{equation}
\label{keff-sub}
D_{\mathrm{eff}}=2\int_0^{\infty}K(s)\langle\xi^2_H\rangle_sds.    
\end{equation}
Analytically determining the exact diffusion coefficient $D_\mathrm{eff}$ is
not feasible but it was proved that they possess a finite value for $H<1/2$
\cite{wang2020a}.\\

\paragraph*{\textbf{MN-FBM-DD:}} The MSD of the MN-FBM-DD (\ref{model-b}) can
be expressed as
\begin{eqnarray}
\label{mn-fbm-msd}
\nonumber
\langle x^2(t)\rangle_{\mathrm{MN}}&=&2V_H\langle D\rangle\Bigg\{\int_0^t(t-s)^{
2H-1 }ds\\
&&\hspace{-1.8cm}+\int_{-\infty}^0 \left[(t-s)^{H-1 / 2}-(-s)^{H-1 / 2}\right]^2ds\Bigg\}.
\end{eqnarray}
The variable transform $z=-s/t$ in Eq.~(\ref{mn-fbm-msd}) then yields
\begin{eqnarray}
\nonumber
\langle x^2(t)\rangle_{\mathrm{MN}}&=&2V_H\langle D\rangle t^{2H}\Bigg\{\int_0
^1\left(1-z\right)^{2H-1}dz\\
\nonumber
&&+\int_0^{+\infty}\left[(1+z)^{H-1/2}-z^{H-1/2}\right]^2dz\Bigg\}\\
&=&2\langle D\rangle t^{2H}.
\label{mn-fbm-msd2}
\end{eqnarray}

\vspace*{0.2cm}

\paragraph*{\textbf{RL-FBM-DD:}} Similarly, we obtain the MSD of the RL-FBM-DD
model (\ref{model-c}),
\begin{eqnarray}
\nonumber
\langle x^2(t)\rangle_{\mathrm{RL}}&=&4H\langle D\rangle\int_0^t(t-s)^{2H-1}ds\\
&=&2\langle D\rangle t^{2H}.
\label{rl-fbm-msd}
\end{eqnarray}
which is identical to result (\ref{mn-fbm-msd2}) for MN-FBM-DD.\\ 

\begin{figure}
(a)\includegraphics[width=0.9\linewidth]{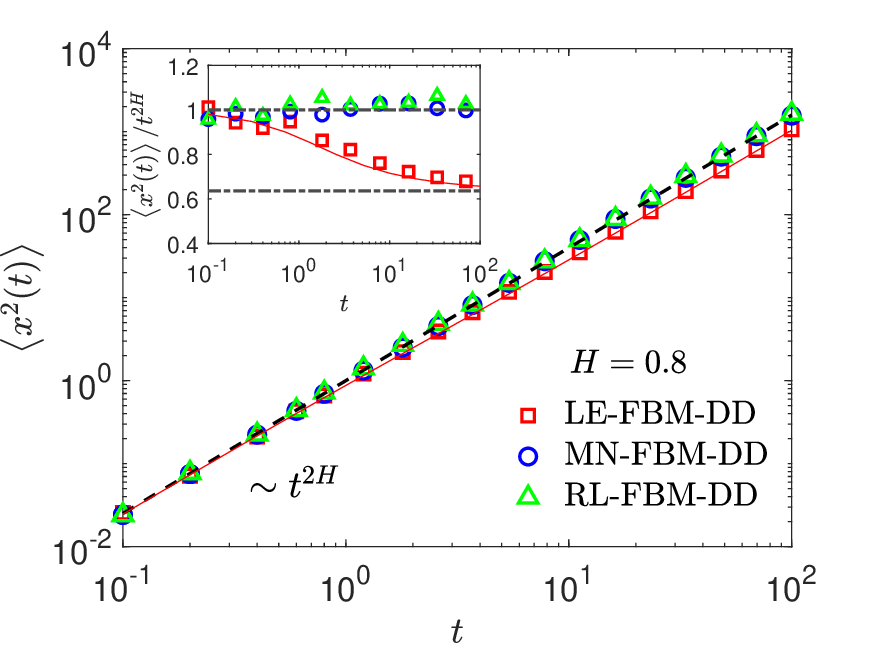}
(b)\includegraphics[width=0.9\linewidth]{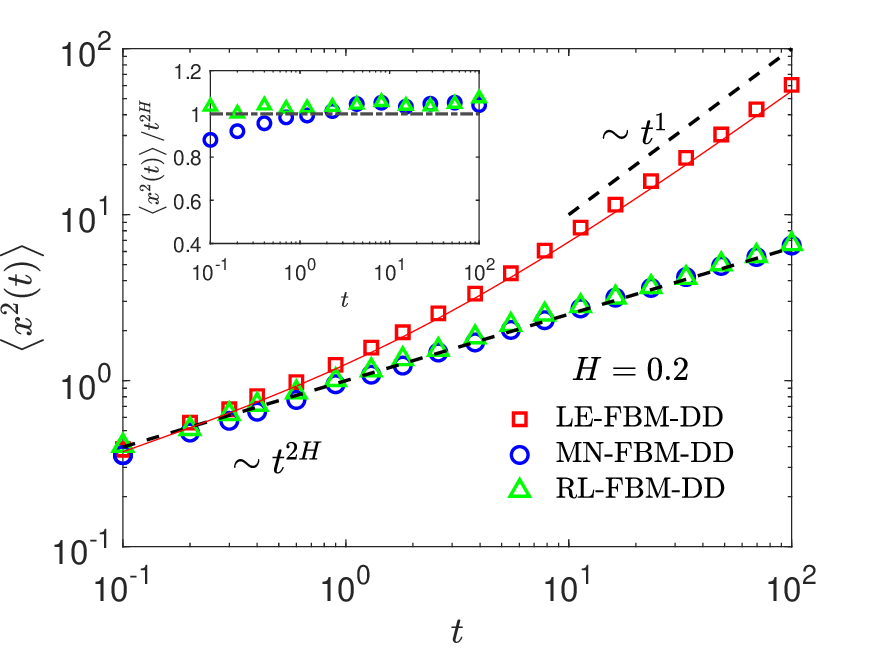}
\caption{Simulations (circles, triangles, rectangles) for the MSD of the three
DD-generalized FBM models with (a) $H=0.8$ and (b) $H=0.2$. The numerical
evaluation of the MSD (\ref{le-fbm-msd}) of the LE-FBM-DD model (\ref{model-a})
is depicted by the red solid curve. The generalized diffusion coefficients are
displayed in the inset. In panel (a), the MSD of all models show the same scaling,
and the generalized diffusion coefficient $\langle x^2(t)\rangle/t^{2H}$ of the
LE-FBM-DD has a crossover from $2\langle D\rangle=1$ to $2K_{\mathrm{eff}}=2/\pi$,
represented by dash-dotted lines. In panel (b), the MSD of the LE-FBM-DD model
switches from $2\langle D\rangle t^{2H}$ at short times to normal diffusion,
$2D_{\mathrm{eff}}t$ at long times, while the MSD of the MN-FBM-DD and RL-FBM-DD
models scale $\simeq t^{2H}$ at all times. Simulation parameters: $\tau=1$,
$\sigma=1$, $dt=0.1$, and $T=100$. These parameters are consistently kept
across all figures, unless stated otherwise.}
\label{fig1}
\end{figure}

Although the classic FBM based on the three representations analyzed here have
the same MSD, the DD-generalized FBM models exhibit significantly different
behaviors: the MSDs of the MN-FBM-DD and RL-FBM-DD models continuously evolve
with a power-law scaling of $t^{2H}$ for all Hurst exponents, with an effective
generalized diffusion coefficient equal to the mean diffusivity; in contrast,
for the LE-FBM-DD model the MSD exhibits a crossover from the short-time
scaling $\simeq t^{2H}$ to the long-time behavior $\simeq t$ for subdiffusion.
The simulations of the MSD for all three models are presented in Fig.~\ref{fig1}
and show excellent agreement with the theory results.

\subsection{MSI}

\paragraph*{\textbf{LE-FBM-DD:}} The MSI of the LE-FBM-DD (\ref{model-a}) model,
\begin{equation}
\label{le-fbm-dd-msi}
\langle x^2_\Delta(t)\rangle_\mathrm{LE}=4\int_0^\Delta(\Delta-s)K(s)
\langle\xi_H^2\rangle_s ds,
\end{equation}
has the same expression as the MSD~(\ref{le-fbm-msd}) and is stationary, i.e.,
solely depending on the lag time $\Delta$. At short times $\Delta\ll\tau$ the
MSD behaves as
\begin{equation}
\label{le-fbm-dd-msi-st}
\langle x^2_\Delta(t)\rangle\sim2\langle D\rangle\Delta^{2H}.
\end{equation}
At long times we find
\begin{equation}
\label{le-fbm-dd-msi-lt}
\langle x_\Delta^2(t)\rangle_\mathrm{LE}=\left\{\begin{array}{ll}2D_{\mathrm{
eff}}\Delta, & H<1/2\\
2K_{\mathrm{eff}}\Delta^{2H}, & H>1/2\end{array}.\right.
\end{equation}
The effective diffusion coefficients $K_{\mathrm{eff}}$ for superdiffusive and
$D_{\mathrm{eff}}$ for subdiffusive $H$ are given by expressions (\ref{K2}) and
(\ref{keff-sub}), respectively.\\

\paragraph*{\textbf{MN-FBM-DD:}} The MSI of the MN-FBM-DD (\ref{model-b}) model
reads
\begin{eqnarray}
\nonumber
\langle x_\Delta^2(t)\rangle_\mathrm{MN}&=&2V_H\langle D\rangle\Bigg\{\int_t^{
t+\Delta}(t+\Delta-s)^{2H-1}ds\\
&&\hspace*{-2.2cm}
+\int_{-\infty}^t\left[(t+\Delta-s)^{H-1/2}-(t-s)^{H-1/2}\right]^2ds\Bigg\}.
\end{eqnarray}
With the variable transform $z=s-t$, the same expression as (\ref{mn-fbm-msd2})
for the MSD of MN-FBM-DD can be obtained,
\begin{eqnarray}
\nonumber
\langle x_\Delta^2(t)\rangle_\mathrm{MN}&=&2V_H\langle D\rangle\Bigg\{\int_0^{
\Delta}(\Delta-z)^{2H-1}dz\\
&&\hspace*{-1.8cm}
+\int_{-\infty}^0\left[(\Delta-z)^{H-1/2}-(-z)^{H-1/2}\right]^2dz\Bigg\},
\end{eqnarray}
which leads to the stationary MSI
\begin{equation}
\label{mn-fbm-dd-msi}
\langle x_\Delta^2(t)\rangle_\mathrm{MN}=2\langle D\rangle \Delta^{2H}.
\end{equation}

\vspace*{0.2cm}

\paragraph*{\textbf{RL-FBM-DD:}} The MSI of RL-FBM (\ref{rl-fbm-msi}) was
analyzed in \cite{lim2001,wei2025}. Here we obtain the MSI of the RL-FBM-DD
model in the form
\begin{equation}
\label{rl-fbm-dd-msi}
\langle x_\Delta^2(t)\rangle_\mathrm{RL}=4H\langle D\rangle\Delta^{2H}
\Bigg\{I_H\left(\frac{t}{\Delta}\right)+\frac{1}{2H}\Bigg\}.
\end{equation}
At short times $t\ll\Delta$, the MSI of RL-FBM-DD is identical to the MSD
(\ref{rl-fbm-msd}),
\begin{eqnarray}
\label{rl-fbm-msi-st}
\langle x_\Delta^2(t)\rangle_\mathrm{RL}\sim 2\langle D\rangle\Delta^{2H}.
\end{eqnarray}
At long times $t\gg\Delta$, the MSI reads
\begin{eqnarray}
\label{rl-fbm-msi-lt}
\langle x_\Delta^2(t)\rangle_\mathrm{RL}\sim\frac{2\langle D\rangle\Gamma(H+1
/2)^2}{\Gamma(2H)\sin(\pi H)}\Delta^{2H}.
\end{eqnarray}

\vspace*{0.2cm}

\begin{figure}
(a)\includegraphics[width=0.9\linewidth]{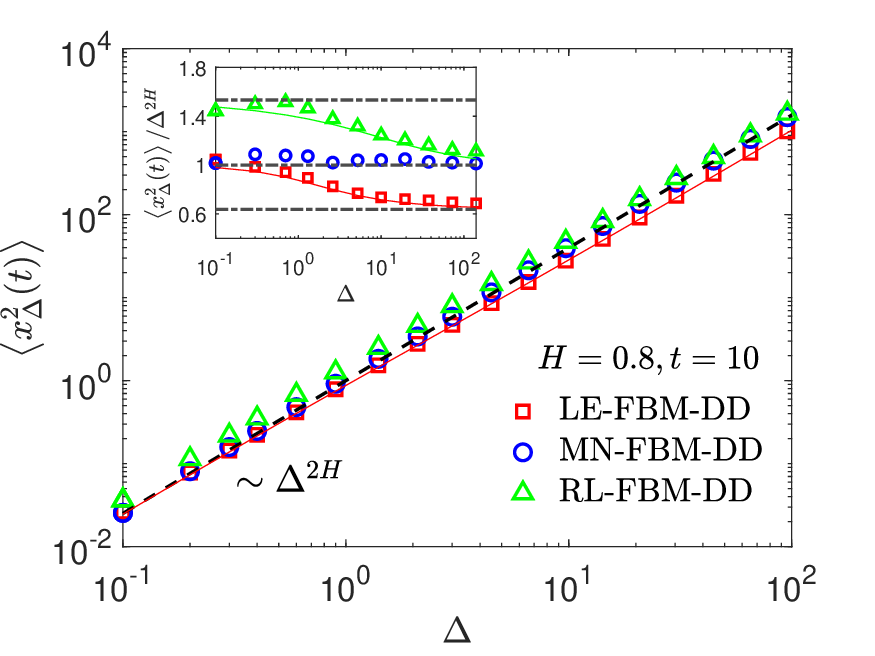}
(b)\includegraphics[width=0.9\linewidth]{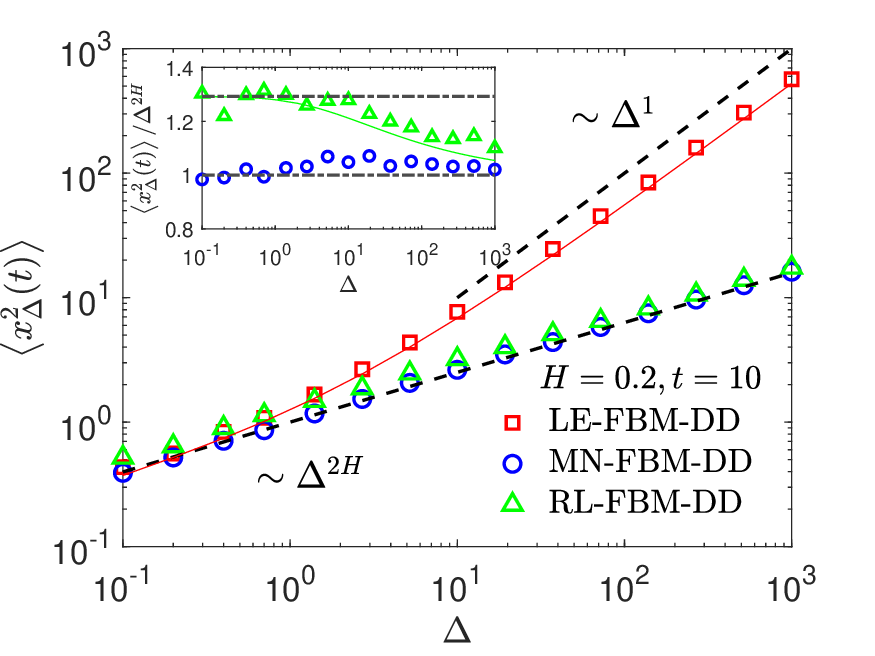}
\caption{Simulations (circles, triangles, rectangles) for the MSIs with
starting time $t=10$ of the three DD-generalized FBM models with (a) $H=0.8$
and (b) $H=0.2$. The numerical evaluations of the MSIs (\ref{le-fbm-dd-msi})
and (\ref{rl-fbm-dd-msi}) of LE-FBM-DD (\ref{model-a}) and RL-FBM-DD
(\ref{model-c}) are represented by red and green curves, respectively. The
generalized diffusion coefficients $\langle x^2(t)\rangle/t^{2H}$ are displayed
in the inset. In panel (a) for the superdiffusive case the MSIs of all models
have the same scaling; the LE-FBM-DD-MSI has a crossover behavior in the
prefactor of the MSI from $2\langle D\rangle=1$ to $2K_{\mathrm{eff}}=2/\pi$,
while for RL-FBM-DD it crosses over from $\frac{2\langle D\rangle\Gamma(H+1/2)
^2}{\Gamma(2H)\sin(\pi H)}$ to $2\langle D\rangle=1$. In panel (b) for the
subdiffusive case, the MSI of the LE-FBM-DD switches from $2\langle D\rangle
\Delta^{2H}$ at short times to normal diffusion $2D_{\mathrm{eff}}\Delta$ at
long times.}
\label{fig2}
\end{figure}

From the MSIs of MN-FBM-DD and RL-FBM-DD, Eqs.~(\ref{mn-fbm-dd-msi}) and
(\ref{rl-fbm-dd-msi}), respectively, we find that the random diffusivity
for MN-FBM-DD and RL-FBM-DD effects an effective diffusion coefficient.  
A crossover behavior similar to that observed in the MSD also appears in
the MSI of the LE-FBM-DD model given by (\ref{le-fbm-dd-msi}). We also note
that the MSI is stationary in the LE-FBM-DD and MN-FBM-DD models but
nonstationary in the RL-FBM-DD model. Figure \ref{fig2} compares the
results from simulations with the theoretical results for the MSI.

\subsection{ACVF}
 
\paragraph*{\textbf{LE-FBM-DD:}} The stationary ACVF of the LE-FBM-DD model
(\ref{model-a}) is given by
\begin{eqnarray}
\nonumber
C^{\delta}_\mathrm{LE}(t,\Delta)&=&2\left<\sqrt{D(t)}\sqrt{D(t+\Delta)}\right>
\langle\xi_H(t)\xi_H(t+\Delta)\rangle\\
&=&2K(\Delta)\langle\xi_H^2\rangle_\Delta.
\label{le-fbm-dd-acf}
\end{eqnarray}
The correlation $K(\Delta)$ in Eq.~(\ref{eq-k_delta}) can be expanded at large
lag times $\Delta\gg\tau$ up to second order, leading to
\begin{eqnarray}
K(\Delta)\sim\sigma^2\tau\left(\pi^{-1}+e^{-2\Delta/\tau}\right),    
\end{eqnarray}
from which the ACVF reads
\begin{eqnarray}
C^{\delta}_\mathrm{LE}(\Delta)\sim2\sigma^2\tau\Big[\pi^{-1}\langle \xi_H^2
\rangle_\Delta+e^{-2\Delta/\tau}\langle\xi_H^2\rangle_\Delta\Big].
\end{eqnarray}
This expression elucidates nicely the a priori unexpected crossover behavior
in the MSD: The first part, proportional to the correlations of the fractional
Gaussian noise, leads to anomalous diffusion characterized by the scaling
$\simeq t^{2H}$ in the MSD at long times whereas the second part, governed by
the truncated power-law noise correlation, contributes the normal-diffusive
scaling $\simeq t^1$ at long times \cite{metz2018}. Especially, when $H<1/2$,
the linear MSD component dominates.\\

\paragraph*{\textbf{MN-FBM-DD:}} To obtain the ACVF of the MN-FBM-DD model
(\ref{model-b}), the correlation of the displacement must be considered,
given by
\begin{equation}
\langle x_\mathrm{MN}(t+\Delta)x_\mathrm{MN}(t)\rangle=\langle D\rangle\big((t
+\Delta)^{2H}+t^{2H}-\Delta^{2H}\big).
\end{equation} 
Substituting the correlation of the displacement into expression (\ref{eq-acf}),
one obtains the stationary ACVF
\begin{equation}
\label{mn-fbm-dd-acf}
C^{\delta}_\mathrm{MN}(t,\Delta)= 2\langle D\rangle\langle\xi_H^2\rangle_\Delta.
\end{equation}
In comparison with the ACVF of MN-FBM (\ref{acf-mn-fbm}), the ACVF of the
MN-FBM-DD model directly reflects an effective diffusion coefficient equal
to the mean diffusivity.\\

\paragraph*{\textbf{RL-FBM-DD:}} Analogously, the ACVF of the RL-FBM-DD model
(\ref{model-c}) turns out to be equal to that of RL-FBM with an effective
diffusion coefficient,
\begin{eqnarray}
\nonumber
C^{\delta}_\mathrm{RL}(t,\Delta)=2\langle D\rangle H(2H-1)\hspace*{3.6cm}\\
\nonumber
\times
\Big\{\frac{3-2H}{2}\Delta^{2H-1}\int_0^{t/\Delta}q^{H-1/2}(1+q)^{H-5/2}dq\\
+\delta^{-1}\Delta^{2H-1}\int_{t/\Delta}^{(t+\delta)/\Delta}q^{H-1/2}(1+q)^{
H-3/2}dq\Big\}.
\label{newacvf}
\end{eqnarray}
Specifically, when $t=0$, the approximate ACVF with $\Delta\gg\delta$ of
RL-FBM-DD is
\begin{equation}
\label{rl-fbm-dd-acf-a}
C^{\delta}_\mathrm{RL}(t,\Delta)\sim\frac{4\langle D\rangle H(2H-1)\delta^{H
-1/2}}{2H+1}\Delta^{H-3/2}.
\end{equation}
When $t\to\infty$, we find
\begin{equation}
\label{rl-fbm-dd-acf-b}
C^{\delta}_\mathrm{RL}(t,\Delta)\sim\frac{2\langle D\rangle H(2H-1)\Gamma(H+1
/2)^2}{\Gamma(2H)\mathrm{sin}(\pi H)}\Delta^{2H-2}.
\end{equation} 

\vspace*{-0.2cm}

\begin{figure}
(a)\includegraphics[width=0.9\linewidth]{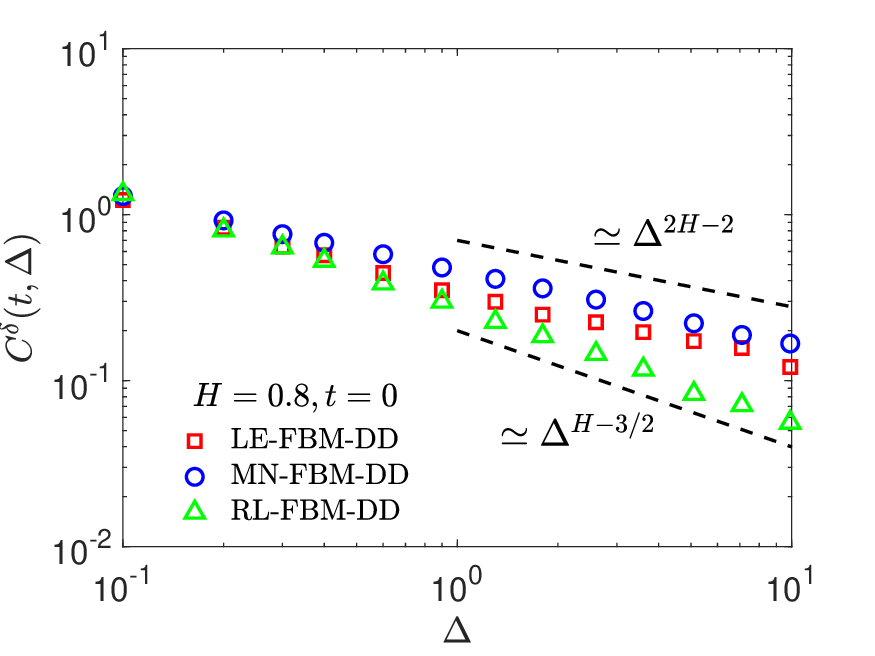}
(b)\includegraphics[width=0.9\linewidth]{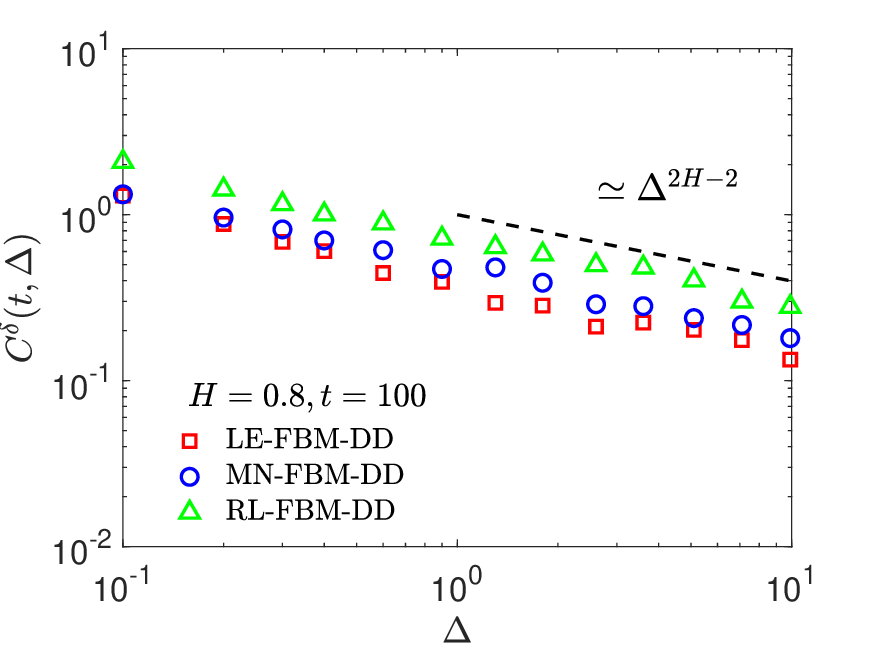}
\caption{Simulations (circles, triangles, rectangles) for the ACVF with
$H=0.8$ and starting times (a) $t=0$ and (b) $t=100$ for the three
DD-generalized FBM models. In panel (a), the ACVF of LE-FBM-DD and MN-FBM-DD
decay as $\simeq\Delta^{2H-2}$, while the ACVF of the RL-FBM-DD model has the
scaling $\simeq\Delta^{H-3/2}$ when $t=0$. In panel (b), the ACVF of all
models have the same scaling $\simeq\Delta^{2H-2}$ at long times, $t=100$.}
\label{fig3}
\end{figure}

To conclude this part, the correlations (\ref{eq-k_delta}) of the random
diffusivity emerges in the ACVF (\ref{le-fbm-dd-acf}) of LE-FBM-DD, effecting
the unexpected crossover behavior in the MSD and MSI. In contrast, the ACVF
reflects the effective diffusion coefficient $\langle D\rangle$ for MN-FBM-DD
and RL-FBM-DD, Eqs.~(\ref{mn-fbm-dd-acf}), (\ref{rl-fbm-dd-acf-a}), and
(\ref{rl-fbm-dd-acf-b}). Moreover, the ACVF is stationary for both LE-FBM-DD
and MN-FBM-DD models, while it is nonstationary in the RL-FBM-DD model. The
simulations for the ACVF are shown in Fig.~\ref{fig3}.

\subsection{PDF}

Given that at short times $t\ll\tau$ the diffusivity following the
Ornstein-Uhlenbeck dynamics changes little over time, single trajectories of
all the three models behave as the corresponding FBM with constant diffusivity,
and the PDFs can be described by a superstatistical approach \cite{beck2003},
i.e., can be obtained as the average of a single Gaussian, with a given
diffusivity, over the stationary diffusivity distribution. Using the same
technique as in \cite{chec2017,wang2020a} the PDF of all three models at
short times yields in the form
\begin{equation}
P(x,t)=\frac{1}{\pi\sqrt{\mathcal{M}(t)_{\mathrm{ST}}}}K_0\left(\frac{|x|}{
\sqrt{\mathcal{M}(t)_{\mathrm{ST}}}}\right),
\end{equation}
where $\mathcal{M}(t)_{\mathrm{ST}}$ is the MSD of the three DD-generalized FBM
models (\ref{le-fbm-msd}), (\ref{mn-fbm-msd2}), and (\ref{rl-fbm-msd}) at short
times. Moreover, $K_0$ denotes the modified Bessel function of the second kind.
In particular, for the relevant large displacements ensuring $z=x/\sqrt{\mathcal{
M}(t)_{\mathrm{ST}}}\gg1$, the Bessel function has the expansion $K_0\sim\sqrt{
\pi/(2z)}e^{-z}$, and thus the PDF can be approximated as
\begin{equation}
\label{pdf-st}
P(x,t)\sim\frac{1}{\sqrt{2\pi|x|\mathcal{M}(t)_{\mathrm{ST}}^{1/2}}}\exp\left(
-\frac{|x|}{\sqrt{\mathcal{M}(t)_{\mathrm{ST}}}}\right).
\end{equation}

\begin{figure}
(a)\includegraphics[width=0.9\linewidth]{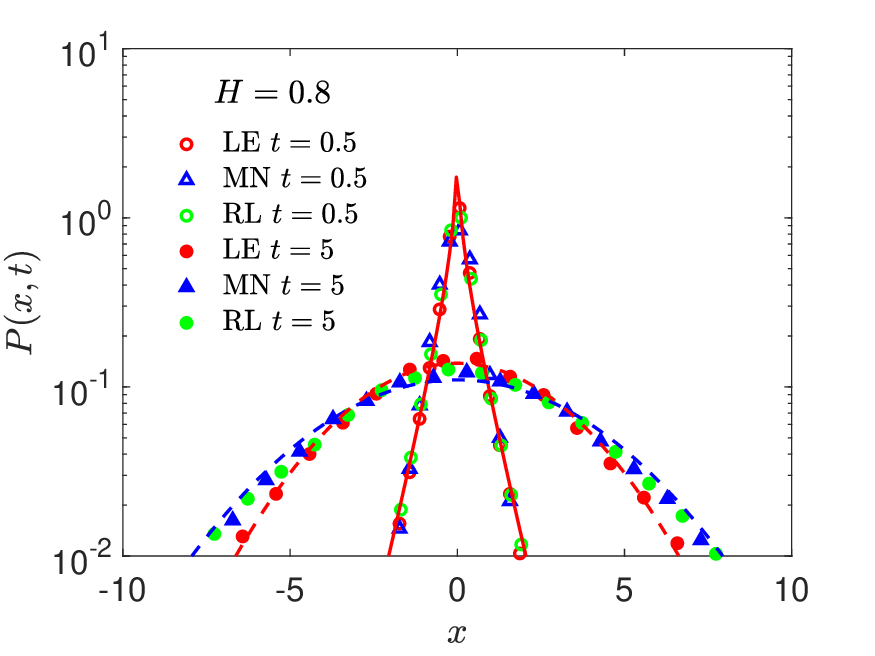}
(b)\includegraphics[width=0.9\linewidth]{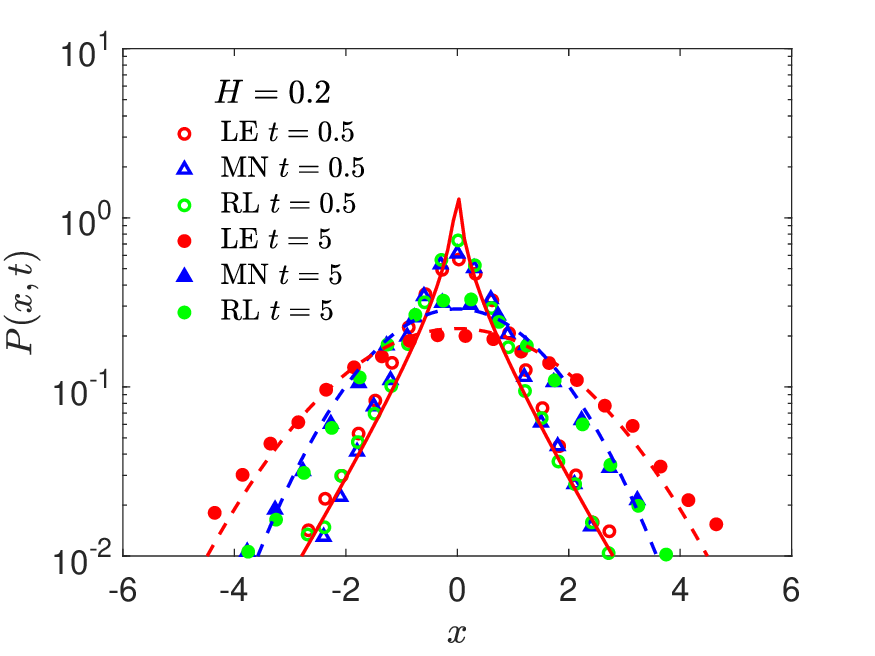}
\caption{Simulations (circles, triangles, rectangles) for the PDF of three
DD-generalized FBM models with (a) $H=0.8$ and (b) $H=0.2$. The theoretical
result (\ref{pdf-st}) and (\ref{pdf-lt}) for the PDF at short and long times,
respectively, are represented by the colored solid and dashed curves.}
\label{fig4}
\end{figure}

At long times, the Gaussian limit is recovered due to the central limit theorem
for all three models,
\begin{eqnarray}
\label{pdf-lt}
P(x,t)\sim\frac{1}{\sqrt{2\pi\mathcal{M}(t)_{\mathrm{LT}}}}\exp\left(-\frac{
x^2}{2\mathcal{M}(t)_{\mathrm{LT}}}\right).
\end{eqnarray}
where $\mathcal{M}(t)_{\mathrm{LT}}$ is the MSD of the three DD-generalized FBM
models (\ref{le-fbm-msd}), (\ref{mn-fbm-msd2}), and (\ref{rl-fbm-msd}). The
comparison of simulations and theoretical results for the PDFs are shown in
Fig.~\ref{fig4}.

\section{Discussion and conclusions}

We examined the statistical properties of the DD-generalizations based on
the three representations of FBM: LE-FBM, MN-FBM, and RL-FBM. The main
results are summarized in Tab.\ref{tab}. With constant coefficients both
LE-FBM and MN-FBM are equivalent, with stationary increments, whereas
RL-FBM has nonstationary increments. However, when introducing the DD
dynamics, the effected dynamics of the DD-generalized FBM processes differ
significantly. The MN-FBM-DD and RL-FBM-DD models share the same MSD, MSI,
and ACVF with their corresponding FBM models, scaled by an effective
diffusivity equal to the mean value. In contrast, the LE-FBM-DD model is
influenced more severely by the diffusivity correlations, leading to a
priori unexpected crossovers in the scaling behaviors of the MSD and MSI.
Additionally, both MN-FBM-DD and LE-FBM-DD exhibit stationary properties,
whereas RL-FBM-DD does not. All DD-generalized FBM models demonstrate a
crossover in the PDF, crossing over from a short-time non-Gaussian
to a long-time Gaussian shape.

The crossover behavior in the scaling of both MSD and MSI observed for the
LE-FBM-DD model can be generally predicted by the correlation of the DD
dynamics following the squared OU process,
\begin{eqnarray}
\nonumber
K(\Delta=|t_1-t_2|)&=&\langle\sqrt{D(t_1)}\sqrt{D(t_2)}\rangle\\
&\sim&c_1+c_2e^{-\Delta/\tau},
\end{eqnarray}
where $c_1,c_2>0$. A similar behavior was also observed when the random
diffusivity follows a Markovian switching behavior ("two-state model")
between the two values ${D_1,D_2}$ with rate $k_1$ and $k_2$ \cite{wang2020a}.
This model is also in equilibrium and has correlations with $c_1,c_2>0$.
Moreover, for other protocols of the diffusivity, no crossover behavior will
be observed when $c_1>0$ and $c_2=0$, e.g., the diffusivity is considered to
be a random variable, while the crossover behavior from anomalous to normal
diffusion occurs for all $H$ when $c_1=0$ and $c_2>0$, e.g., in the
Tyagi-Cherayil model \cite{tyagi} defined by the Langevin dynamics $dx/dt=
\sqrt{2}Y(t)\xi_H(t)$ \cite{wang2020a}.

Over the recent years, extensive data from modern single particle tracking
and supercomputing studies in complex systems, e.g., bio-relevant systems,
demonstrated that the observed dynamics is often more complicated and can
no longer be described by individual (anomalous) diffusion processes.
Classical statistical observables \cite{weis2013,weig2011,siko2017,vester},
Bayesian statistics \cite{kevin,michael,thap2018}, or deep learning-based
analyses \cite{gorka,gork2021,seck2023,manzo2021,seck2022,szwa2019,gran2019,
han2020} by-now provide quite reliable means to decipher the (anomalous)
diffusive process(es) underlying the data. To provide an ever better body
of models for such analyses, the DD-generalized FBM models discussed here
should be added to these algorithms.

We anticipate that our results will prompt research on a deeper understanding
of how heterogeneity influences diffusion transport phenomena. Several
intriguing issues warrant further exploration. For instance, FBM with DD
considered to be non-Markovian state-switching was recently studied in
\cite{balc25}, and the system remains non-Gaussian across all time scales,
underscoring the role of heavy-tailed distributions in shaping the statistical
properties of diffusion processes. Similar persistent non-Gaussian behavior
was also found for particles with DD in a confined harmonic potential
\cite{yann22}. Recent single-particle-tracking experiments revealed that
intracellular transport of endo- and exo-genous tracers of various sizes is
often not only anomalous, but also heterogeneous in time and space. This
implies that a single diffusion exponent of standard anomalous-diffusion
models is insufficient to describe the underlying physical phenomena. The
DD-generalized FBM model, for which the anomalous diffusion exponent depends
on time or space deserves to be investigated. Additional aspects, such as
the first passage dynamics \cite{vitt2024a,vitt2024b} and resetting dynamics
\cite{evan11,pal17} should also be explored. Finally, another intriguing
direction of the research could be to study universal singularities of the
three models with random diffusion exponents along the concepts developed
in \cite{stella1,stella2}.

\begin{table*}
\centering
\begin{ruledtabular}
\begin{tabular}{|c|c|c|c|}
& \textbf{LE-FBM-DD} & \textbf{MN-FBM-DD} & \textbf{RL-FBM-DD} \\
\hline
$\langle x(t)^2\rangle$ & \parbox[t]{6cm}{Short time: $\sim 2\langle D \rangle
t^{2H}$, Eq.~(\ref{le-fbm-dd-msd-st})\\
Long time ($H>1/2$): $\sim2K_{\mathrm{eff}}t^{2H}$,
Eq.~(\ref{le-fbm-dd-msd-lt-1})\\
Long time ($H<1/2$): $\sim2D_{\mathrm{eff}}t$, Eq.~(\ref{le-fbm-dd-msd-lt-2})} 
& $2\langle D\rangle t^{2H}$, Eq.~(\ref{mn-fbm-msd2})
& $2\langle D\rangle t^{2H}$, Eq.~(\ref{rl-fbm-msd})\\
\hline
{$\langle x^2_\Delta(t)\rangle$}
& \parbox[t]{6cm}{Stationary:\\
Short lag time: $\sim2\langle D\rangle\Delta^{2H}$, Eq.~(\ref{le-fbm-dd-msi-st})\\
Long lag time ($H>1/2$): $\sim2K_{\mathrm{eff}}\Delta^{2H}$,
Eq.~(\ref{le-fbm-dd-msi-lt})\\
Long lag time ($H<1/2$): $\sim2D_{\mathrm{eff}}\Delta$,
Eq.~(\ref{le-fbm-dd-msi-lt})} 
& \parbox[t]{3.64cm}{Stationary:\\
$2\langle D\rangle\Delta^{2H}$, Eq.~(\ref{mn-fbm-dd-msi}) }
& \parbox[t]{6cm}{Nonstationary: $4H\langle D\rangle\left[I_H\left(\frac{t}{
\Delta}\right)+\frac{1}{2H}\right]\Delta^{2H}$, Eq.~(\ref{rl-fbm-dd-msi})\\
Short time: $\sim2\langle D\rangle\Delta^{2H}$, Eq.~(\ref{rl-fbm-msi-st})\\
Long time: $\sim\frac{2\langle D\rangle\Gamma(H+1/2)^2}{\Gamma(2H)\sin(\pi
H)}\Delta^{2H}$, Eq.~(\ref{rl-fbm-msi-lt})}\\
\hline
{$C^{\delta}(t,\Delta)$}
& \parbox[t]{6cm}{Stationary:\\
$2K(\Delta)\langle\xi_H^2\rangle_\Delta$, Eq.~(\ref{le-fbm-dd-acf})} 
& \parbox[t]{3.6cm}{Stationary:\\
$2\langle D \rangle\langle\xi_H^2\rangle_\Delta$, Eq.~(\ref{mn-fbm-dd-acf})}
& \parbox[t]{6cm}{Nonstationary: Eq.~(\ref{newacvf})\\
$t=0$: $\sim\frac{4\langle D\rangle H(2H-1)\delta^{H-1/2}}{2H+1}\Delta^{H-3/2}$,
Eq.~(\ref{rl-fbm-dd-acf-a})\\
$t\to\infty$: $\sim\frac{2\langle D\rangle H(2H-1)\Gamma\left(H+1/2\right)^2}{
\Gamma(2H)\mathrm{sin}(\pi H)}\Delta^{2H-2}$, Eq.~(\ref{rl-fbm-dd-acf-b})}\\
\end{tabular}
\end{ruledtabular}
\caption{Comparison of the statistical properties of the three FBM-generalized DD models: LE-FBM-DD, MN-FBM-DD, and RL-FBM-DD.}
\label{tab}
\end{table*}

\begin{acknowledgments}
R.M. acknowledges financial support from the German Science Foundation (DFG,
Grant ME 1535/13-1 and ME 1535/22-1) and NSF-BMBF CRCNS (Grant 2112862/STAXS).
A.V.C. acknowledges BMBF Project PLASMA-SPIN Energy (Grant 01DK2406).
\end{acknowledgments}


\begin{thebibliography}{99}

\bibitem{brow1828} R. Brown, A brief account of microscopical
observations made in the months of June, July and August 1827,
on the particles contained in the pollen of plants; and on the
general existence of active molecules in organic and inorganic bodies,
\href{https://doi.org/10.1080/14786442808674769}{Philos. Mag. \textbf{4},
161 (1828)}.

\bibitem{eins1905} A. Einstein, {\"{U}}ber die von der
molekularkinetischen Theorie der W{\"{a}}rme geforderte Bewegung
von in ruhenden Fl{\"{u}}ssigkeiten suspendierten Teilchen,
\href{https://doi.org/10.1002/andp.19053220806}{Ann. Phys. \textbf{322},
549 (1905)}.

\bibitem{suth1905} W. Sutherland, A dynamical theory of
diffusion for nonelectrolytes and the molecular mass of albumin,
\href{https://doi.org/10.1080/14786440509463331}{Philos. Mag. \textbf{9},
781 (1905)}.

\bibitem{smol1906} M. von Smoluchowski, Zur kinetischen Theorie
der Brownschen Molekularbewegung und der Suspensionen, \href{
https://doi.org/10.1002/andp.19063261405}{ Ann. Phys. \textbf{326}, 756
(1906)}.

\bibitem{lang1908} P. Langevin, Sur la th{\'e}orie du mouvement brownien,
C. R. Acad. Sci. (Paris) \textbf{146}, 530 (1908).

\bibitem{perrin} J. Perrin, Les atomes (F{\'e}lix Alcan, Paris, 1913);
English version: J. Perrin and D. L. Hammick (translator), Atoms (Kessinger
Publishing, Whitefish, MT).

\bibitem{metz2014} R. Metzler, J-H. Jeon, A. Cherstvy, and E. Barkai,
Anomalous diffusion models and their properties: non-stationarity,
non-ergodicity, and ageing at the centenary of single particle tracking,
\href{https://doi.org/10.1039/C4CP03465A}{Phys. Chem. Chem. Phys. \textbf{16},
24128 (2014)}.

\bibitem{weis2004} M. Weiss, M. Elsner, F. Kartberg, and T. Nilsson,
Anomalous subdiffusion is a measure for cytoplasmic crowding in living cells,
\href{https://doi.org/10.1529/biophysj.104.044263}{Biophys J. \textbf{87},
3518 (2004)}.

\bibitem{gold2006} I. Golding and E. C. Cox, Physical nature of bacterial
cytoplasm, \href{https://doi.org/10.1103/PhysRevLett.96.098102}{Phys. Rev.
Lett. \textbf{96}, 098102 (2006)}.

\bibitem{jeon2011} J.-H. Jeon, V. Tejedor, S. Burov, E. Barkai,
C. Selhuber-Unkel, K. Berg-S{\o}rensen, L. Oddershede, and R. Metzler, In
Vivo Anomalous diffusion and weak ergodicity breaking of lipid granules,
\href{https://doi.org/10.1103/PhysRevLett.106.048103}{Phys. Rev. Lett.
\textbf{106}, 048103 (2011)}.

\bibitem{beta2024} R. Gro{\ss}mann, L. S. Bort, T. Moldenhawer,
M. Stange, S. S. Panah, R. Metzler, and C. Beta, Non-Gaussian
displacements in active transport on a carpet of motile cells,
\href{https://doi.org/10.1103/PhysRevLett.132.088301}{Phys. Rev. Lett.
\textbf{132}, 088301 (2024)}.

\bibitem{jeon2016} J. -H. Jeon, M. Javanainen, H. Martinez-Seara, R. Metzler,
and I. Vattulainen, Protein Crowding in lipid bilayers gives rise to
non-Gaussian anomalous lateral diffusion of phospholipids and proteins,
\href{https://doi.org/10.1103/physrevx.6.021006}{Phys.  Rev. X \textbf{6},
021006 (2016)}.

\bibitem{gode2014} A. Godec, M. Bauer, and R. Metzler, Collective dynamics
effect transient subdiffusion of inert tracers in flexible gel networks,
\href{https://doi.org/10.1088/1367-2630/16/9/092002}{New J. Phys. \textbf{16},
092002 (2016)}.

\bibitem{akim2011} T. Akimoto, E. Yamamoto, K. Yasuoka,
Y. Hirano, and M. Yasui, Non-Gaussian Fluctuations
Resulting from Power-Law Trapping in a Lipid Bilayer,
\href{https://doi.org/10.1103/PhysRevLett.107.178103}{Phy. Rev. Lett.
\textbf{107}, 178103 (2011)}.

\bibitem{mich2003} F. Michael, M.D. Johnson, Financial market dynamics,
\href{https://doi.org/10.1016/S0378-4371(02)01558-3}{Physica A \textbf{320},
525 (2003)}.

\bibitem{mant2000} R.N. Mantegna, H.E. Stanley, An introduction to
econophysics: correlations and complexity in finance (Cambridge University
Press, Cambridge, 2000).

\bibitem{hump2010} N. E. Humphries et al., Environmental context
explains L{\'e}vy and Brownian movement patterns of marine predators,
\href{https://doi.org/10.1038/nature09116}{Nature \textbf{465}, 1066 (2010)}.

\bibitem{meye2023} P. G. Meyer, A. G. Cherstvy, H. Seckler,
R. Hering, N. Blaum, F. Jeltsch, and R. Metzler,
Directedness, correlations, and daily cycles in springbok
motion: from data via stochastic models to movement prediction,
\href{https://doi.org/10.1103/PhysRevResearch.5.043129}{Phys. Rev. Res.
\textbf{5}, 043129 (2023)}.

\bibitem{vilk2022a} O. Vilk, Y. Orchan, M. Charter, N. Ganot, S. Toledo,
R. Nathan, and M. Assaf, Ergodicity breaking in area-restricted search of
Avian predators, \href{https://doi.org/10.1103/PhysRevX.12.031005}{Phys. Rev. X
\textbf{12}, 031005 (2022)}.

\bibitem{vilk2022b} O. Vilk et al., Unravelling the origins
of anomalous diffusion: From molecules to migrating storks,
\href{https://doi.org/10.1103/PhysRevResearch.4.033055}{Phys. Rev. Res.
\textbf{4}, 033055 (2022)}.

\bibitem{yan2014} A. Lazarian, and H. Yan, Superdiffusion
of cosmic rays: implications for cosmic ray acceleration,
\href{https://doi.org/10.1088/0004-637X/784/1/38}{Astrophys. J. \textbf{784},
38 (2014)}.

\bibitem{brian} B. Berkowitz, A. Cortis, M. Dentz, and H. Scher, Modeling
non-Fickian transport in geological formations as a continuous time random
walk, Rev. Geophys. \textbf{44}, RG2003 (2006).

\bibitem{brian1} A. Rajyaguru, R. Metzler, I. Dror, D. Grolimund, and B.
Berkowitz, Diffusion in porous rock is anomalous, Env. Sci. Technol.
\textbf{58}, 8946 (2024).

\bibitem{wei2022} D. Wei, A. Rubio-Abadal, B. Ye, F. Machado, J. Kemp,
K. Srakaew, S. Hollerith, J. Rui, S. Gopalakrishnan, N. Y. Yao, I. Bloch,
and J. Zeiher, Quantum gas microscopy of Kardar-Parisi-Zhang superdiffusion,
\href{https://doi.org/10.1126/science.abk2397}{Science \textbf{376}, 716
(2022)}.

\bibitem{bouchaud} J.-P. Bouchaud and A. Georges, Anomalous diffusion in
disordered media: Statistical mechanisms, models and physical applications,
Phys. Rep. \textbf{195}, 127 (1990).

\bibitem{igor} I. M. Sokolov, Models of anomalous diffusion in crowded
environments, Soft Matt. \textbf{8}, 9043 (2012).

\bibitem{pt} E. Barkai, Y. Garini, and R. Metzler, Strange kinetics of single
molecules in living cells, Phys. Today \textbf{65(8)}, 29 (2012).

\bibitem{mishura} Y. Mishura, Stochastic calculus for fractional Brownian
motion and related processes (Springer, Berlin, 2008).

\bibitem{sikorski} M. M. Meerschaert and A. Sikorskii, Stochastic models for
fractional calculus, Vol. 43 (Walter de Gruyter, Berlin, 2019).

\bibitem{mont1965} E. W. Montroll and
G. H. Weiss, Random walks on lattices. II,
\href{https://doi.org/10.1063/1.1704269}{J. Math. Phys. \textbf{6}, 167
(1965)}.

\bibitem{klablushle} J. Klafter, A. Blumen, and M. F. Shlesinger, Stochastic
pathway to anomalous diffusion, Phys. Rev. A \textbf{35}, 3081 (1987).

\bibitem{hughes} B. D. Hughes, Random walks and random environments, Vol. 1:
Random walks (Oxford University Press, Oxford, UK, 1995).

\bibitem{report} R. Metzler and J. Klafter, The random walk's guide to
anomalous diffusion: a fractional dynamics approach, Phys. Rep. \textbf{339},
1 (2000).

\bibitem{mandelbrot} J. M. Berger and B. Mandelbrot, A new model for error
clustering in telephone circuits, IBM J. \textbf{July}, 224 (1963).

\bibitem{harvey} H. Scher and E. W. Montroll, Anomaous transit-time dispersion
in amorphous solids, Phys. Rev. B \textbf{12}, 2455 (1975).

\bibitem{kolm1940} A. N. Kolmogorov, Wienersche Spiralen und einige andere
interessante Kurven im Hilbertschen Raum, C. R. (Doklady) Acad. Sci. URSS
(N.S.) \textbf{26}, 115 (1940).

\bibitem{mand1968} B. B. Mandelbrot and J. W. van Ness,
Fractional Brownian motions, fractional noises and applications,
\href{https://doi.org/10.1137/1010093}{SIAM Rev. \textbf{10}, 422 (1968)}.

\bibitem{weis2007} G. Guigas, V. Kalla, and M. Weiss, Probing the
nanoscale viscoelasticity of intracellular fluids in living cells,
\href{https://doi.org/10.1529/biophysj.106.099267}{Biophys. J. \textbf{93},
316 (2007)}.

\bibitem{spak2010} S. C. Weber, A. J. Spakowitz, and J. A. Theriot,  Bacterial
chromosomal loci move subdiffusively through a viscoelastic cytoplasm,
\href{https://doi.org/10.1103/PhysRevLett.104.238102}{Phys. Rev. Lett.
\textbf{104}, 238102 (2010)}.

\bibitem{janu2020}  S. Janu\v{s}onis, N. Detering, R. Metzler,
and T. Vojta, Serotonergic axons as fractional brownian
motion paths: insights into the self-organization of regional
densities, \href{https://doi.org/10.3389/fncom.2020.00056}{
Front. Comp. Neurosci. \textbf{14}, 56 (2020)}.

\bibitem{janu2023} S. Janu\v{s}onis, J. H.  Haiman, R. Metzler,
and T. Vojta, Predicting the distribution of serotonergic axons: a
supercomputing simulation of reflected fractional Brownian motion in a
3D-mouse brain model, \href{ https://doi.org/10.3389/fncom.2023.1189853}{
Front. Comp. Neurosci. \textbf{17}, 1189853 (2023)}.

\bibitem{gath2018} J. Gatheral, T. Jaisson, and M. Rosenbaum, Volatility is
rough, \href{https://dx.doi.org/10.2139/ssrn.2509457}{Quant. Fin. \textbf{18},
933 (2018)}.

\bibitem{guggenberger} T. Guggenberger, G. Pagnini, T. Vojta, and R. Metzler,
Fractional Brownian motion in a finite interval: correlations effect depletion
or accretion zones of particles near boundaries, New J. Phys. \textbf{21},
022002 (2019).

\bibitem{vojta} A. H. O. Wada and T. Vojta, Fractional Brownian motion with
a reflecting wall, Phys. Rev. E \textbf{97}, 020102(R) (2018).

\bibitem{guggenberger1} T. Vojta, S. Halladay, S. Skinner, S. Janu\v{s}onis,
T. Guggenberger, and R. Metzler, Reflected fractional Brownian motion in
one and higher dimensions, Phys. Rev. E \textbf{102}, 032108 (2020).

\bibitem{gugg} T. Guggenberger, A. Chechkin, and R. Metzler, Fractional
Brownian motion in superharmonic potentials and non-Boltzmann stationary
distributions, J. Phys. A \textbf{54}, 29LT01 (2021).

\bibitem{gugg1} T. Guggenberger, A. V. Chechkin, and R. Metzler, Absence
of confinement and non-Boltzmann stationary states of fractional Brownian
motion in shallow external potentials, New J. Phys. \textbf{24}, 073006 (2022).

\bibitem{krap2024} A. Pacheco-Pozo and D. Krapf,
Fractional Brownian motion with fluctuating diffusivities,
\href{https://doi.org/10.1103/PhysRevE.110.014105}{Phys. Rev. E \textbf{110},
014105 (2024)}.

\bibitem{wang2023} W. Wang, M. Balcerek, K. Burnecki, A. Chechkin,
S. Janu{\v{s}}onis, J. {\'S}l{\k{e}}zak, T. Vojta, A. Wy{\l}oma{\'n}ska,
and R. Metzler, Memory-multi-fractional Brownian motion with continuous
correlations, \href{https://doi.org/10.1103/PhysRevResearch.5.L032025}{Phys.
Rev. Res. \textbf{5}, L032025 (2023)}.

\bibitem{balc2023} M. Balcerek, A. Wy{\l}oma{\'n}ska, K. Burnecki, R. Metzler,
and D. Krapf, Modelling intermittent anomalous diffusion with switching
fractional Brownian motion, \href{https://doi.org/10.1088/1367-2630/ad00d7}{New
J. Phys. \textbf{25}, 103031 (2023)}.

\bibitem{jakub} J. {\'S}l\k{e}zak and R. Metzler, Minimal model of diffusion
with time changing Hurst exponent, J. Phys. A \textbf{56}, 35LT01 (2023).

\bibitem{waig2023} T. A. Waigh and N. Korabel, Heterogeneous
anomalous transport in cellular and molecular biology,
\href{https://doi.org/10.1088/1361-6633/ad058f}{Rep. Prog. Phys. \textbf{86},
126601 (2023)}.

\bibitem{wang2012} B. Wang, J. Kuo, S. C. Bae, and
S. Granick, When Brownian diffusion is not Gaussian,
\href{https://doi.org/10.1038/nmat3308}{Nat. Mater. \textbf{11}, 481 (2012)}.

\bibitem{wang2009} B. Wang, S. M. Anthony, S. C. Bae, and S. Granick,
Anomalous yet Brownian, \href{https://doi.org/10.1073/pnas.0903554106}{Proc.
Natl. Acad. Sci. U.S.A. \textbf{106}, 15160 (2009)}.

\bibitem{miot2021} J. M. Miotto, S. Pigolotti, A. V. Chechkin, and
S. Rold{\'a}n-VargasLength, Length scales in Brownian yet non-Gaussian
dynamics, \href{https://doi.org/10.1103/PhysRevX.11.031002}{Phys. Rev. X
\textbf{11}, 031002 (2021)}.

\bibitem{alex2023} A. Alexandre, M. Lavaud, N. Fares, E. Millan, Y. Louyer,
T. Salez, Y. Amarouchene, T. Gu{\'e}rin, and D. S. Dean, Non-Gaussian diffusion
near surfaces, \href{https://doi.org/10.1103/PhysRevLett.130.077101}{Phys.
Rev. Lett. \textbf{130}, 077101 (2023)}.

\bibitem{rusc2022} F. Rusciano, R. Pastore, and F. Greco,
Fickian Non-Gaussian diffusion in glass-forming liquids,
\href{https://doi.org/10.1103/PhysRevLett.128.168001}{Phys. Rev. Lett.
\textbf{128}, 168001 (2022)}.

\bibitem{roic2020} I. Chakraborty and Y. Roichman, Disorder-induced
Fickian yet non-Gaussian diffusion in heterogeneous media,
\href{https://doi.org/10.1103/PhysRevResearch.2.022020}{Phys. Rev. Res.
\textbf{2}, 022020(R) (2020)}.

\bibitem{seno2022} S. Nampoothiri, E. Orlandini , F. Seno, and F. Baldovin,
Brownian non-Gaussian polymer diffusion and queuing theory in the
mean-field limit, \href{https://doi.org/10.1088/1367-2630/ac4924}{New
J. Phys. \textbf{24}, 023003 (2022)}.

\bibitem{chub2014} M.V. Chubynsky and G.W. Slater, Diffusing
Diffusivity: A Model for anomalous, yet Brownian, diffusion,
\href{https://doi.org/10.1103/PhysRevLett.113.098302}{Phys. Rev. Lett.
\textbf{113}, 098302 (2014)}.

\bibitem{chec2017} A.V. Chechkin, F. Seno, R. Metzler, I.M. Sokolov, Brownian
yet non-Gaussian diffusion: from superstatistics to subordination of diffusing
diffusivities, \href{https://doi.org/10.1103/PhysRevX.7.021002}{Phys. Rev. X
\textbf{7}, 021002 (2017)}.

\bibitem{vittoria} V. Sposini, A. V. Chechkin, F. Seno, G. Pagnini, and R.
Metzler, Random diffusivity from stochastic equations: comparison of two
models for Brownian yet non-Gaussian diffusion, New J. Phys., \textbf{20},
043044 (2018).

\bibitem{vittoria1} V. Sposini, D. S. Grebenkov, R. Metzler, G. Oshanin,
and F.  Seno, Universal spectral features of different classes of random
diffusivity processes, New J. Phys. \textbf{22}, 063056 (2020).

\bibitem{tyagi} N. Tyagi and B. J. Cherayil, Non-Gaussian Brownian diffusion
in dynamically disordered thermal environments, J. Phys. Chem. B \textbf{121},
29 (2017).

\bibitem{jain2017} R. Jain and K.L. Sebastian, L{\'e}vy flight
with absorption: a model for diffusing diffusivity with long tails,
\href{https://doi.org/10.1103/PhysRevE.95.032135}{Phys. Rev. E \textbf{95},
032135 (2017)}.

\bibitem{jain2016} R. Jain and K.L. Sebastian, Diffusion in a Crowded,
Rearranging Environment, \href{https://doi.org/10.1021/acs.jpcb.6b01527}{
J. Phys. Chem. B  \textbf{120}, 3988 (2016)}.

\bibitem{spak2017} T. J. Lampo, S. Stylianidou, M. P. Backlund,
P. A. Wiggins, and A. J. Spakowitz, Cytoplasmic RNA-Protein
Particles Exhibit Non-Gaussian Subdiffusive Behavior,
\href{https://doi.org/10.1016/j.bpj.2016.11.3208.}{Biophys. J. \textbf{112},
532 (2017)}.

\bibitem{tong2016} W. He, H. Song, Y. Su, L. Geng, B. J. Ackerson,
H. B. Peng, and P. Tong, Dynamic heterogeneity and non-Gaussian
statistics for acetylcholine receptors on live cell membrane,
\href{https://doi.org/10.1038/ncomms11701}{Nat. Commun. \textbf{7}, 11701
(2016)}.

\bibitem{thapa} S. Thapa, N. Lukat, C. Selhuber-Unkel, A. G. Cherstvy, and R.
Metzler, Transient superdiffusion of polydisperse vacuoles in highly motile
amoeboid cells, J. Chem. Phys. \textbf{150}, 144901 (2019).

\bibitem{beta} A. G. Cherstvy, O. Nagel, C. Beta, and R. Metzler,
Non-Gaussianity, population heterogeneity, and transient superdiffusion in
the spreading dynamics of amoeboid cells, Phys. Chem. Chem. Phys. \textbf{20},
23034 (2018).

\bibitem{amanda} A. D{\'i}ez Fernandez, P. Charchar, A. G. Cherstvy,
R. Metzler, and M. W. Finnis, The diffusion of doxorubicin drug molecules in
silica nanochannels is non-Gaussian and intermittent, Phys. Chem. Chem. Phys.
\textbf{22}, 27955 (2020).

\bibitem{wang2020a} W. Wang, F. Seno, I. M. Sokolov, A. V. Chechkin, and
R. Metzler, Unexpected crossovers in correlated random-diffusivity processes,
\href{https://doi.org/10.1088/1367-2630/aba390}{New J. Phys. \textbf{22},
083041 (2020)}.

\bibitem{wang2020b} W. Wang, A. G. Cherstvy, A. V. Chechkin, S. Thapa,
F. Seno, X. Liu, and R. Metzler, Fractional Brownian motion with random
diffusivity: emerging residual nonergodicity below the correlation
time, \href{http://doi.org/10.1088/1751-8121/aba467}{J. Phys. A:
Math. Theor. \textbf{53}, 474001 (2020).}

\bibitem{agri} L. Luo and M. Yi, Non-Gaussian diffusion in static disordered
media, Phys. Rev. E \textbf{97}, 042122 (2018).

\bibitem{agri1} L. Luo and M. Yi, Quenched trap model on the extreme landscape:
The rise of subdiffusion and non-Gaussian diffusion, Phys. Rev. E \textbf{100},
042136 (2019).

\bibitem{seongyu} S. Park, X. Durang, R. Metzler, and J.-H. Jeon, Fickian yet
non-Gaussian diffusion in an annealed heterogeneous environment; E-print
arXiv:2503.15366.

\bibitem{wei2025} Q. Wei, W. Wang, Y. Tang, R. Metzler, and A. Chechkin,
Fractional Langevin equation far from equilibrium: Riemann-Liouville
fractional Brownian motion, spurious nonergodicity, and aging,
\href{https://doi.org/10.1103/PhysRevE.111.014128}{Phys. Rev. E \textbf{111},
014128}.

\bibitem{kolm1941a} A.N. Kolmogorov, The local structure of turbulence
in incompressible viscous fluid for very large Reynolds numbers,
Dokl. Akad. Nauk. SSSR \textbf{30}, 301 (1941).

\bibitem{kolm1941b} A.N. Kolmogorov, Dissipation of energy in locally
isotropic turbulence, Dokl. Akad. Nauk. SSSR \textbf{32}, 19 (1941).

\bibitem{yagl1953} A. M. Yaglom, and M.S. Pinsker, Random process with
stationary increments of order $n$, Dokl. Akad. Nauk. SSSR \textbf{90}, 731
(1953).

\bibitem{yagl1987} A. M. Yaglom, Correlation Theory of Stationary and Related
Random Functions, Vol. 1, Basic Results, Vol. 2, Supplementary Notes and
References (Springer, New York, 1987).

\bibitem{mari99} D. Marinucci and P. M. Robinson,
Alternative forms of fractional Brownian motion,
\href{https://doi.org/10.1016/S0378-3758(98)00245-6}{J. Statist. Plan.
Infer. \textbf{80}, 111 (1999)}.

\bibitem{balc2022} M. Balcerek, K. Burnecki, S. Thapa, A. Wy{\l}oma{\'n}ska,
and A. V. Chechkin, Fractional Brownian motion with random Hurst
exponent: Accelerating diffusion and persistence transitions,
\href{https://doi.org/10.1063/5.0101913}{ Chaos \textbf{32}, 093114 (2022)}.

\bibitem{metz2018} D. Molina-Garcia, T. Sandev, H. Safdari, G. Pagnini,
A. Chechkin, and R. Metzler, Crossover from anomalous to normal diffusion:
truncated power-law noise correlations and applications to dynamics
in lipid bilayers, \href{https://doi.org/10.1088/1367-2630/aae4b2}{New
J. Phys. \textbf{20},  103027 (2018)}.

\bibitem{levy} P. L{\'e}vy, in Random functions: General theory with special
reference to Laplacian random functions (University of California Press,
Berkeley, 1953).

\bibitem{lim2001} S. C. Lim, Fractional Brownian motion and
multifractional Brownian motion of Riemann-Liouville type,
\href{http://doi.org/10.1088/0305-4470/34/7/306}{ J. Phys. A:
Math. Theor. \textbf{34}, 1301 (2001)}.

\bibitem{hosk1984} J. R. M. Hosking, Modeling persistence
in hydrological time series using fractional differencing,
\href{http://doi.org/10.1029/WR020i012p01898}{Water Resour. Res. \textbf{20},
1898 (1984)}.

\bibitem{diek2004} D. Dieker, Simulation of fractional Brownian motion,
Master's thesis, University of Twente, The Netherlands (2004).

\bibitem{inference} J. F. Coeurjolly, Inf{\'e}rence statistique pour les
mouvements browniens fractionnaires et multifractionnaires, Doctoral
dissertation, Universit{\'e} Joseph-Fourier-Grenoble I, France (2000).

\bibitem{wood1994} A. T. Wood and G. Chan, Simulation
of stationary Gaussian processes in $[0, 1]^d$ ,
\href{http://doi.org/10.2307/1390903}{J. Comput. Graph. Stat. \textbf{3},
409 (1994)}.

\bibitem{pina1994} S. Rambaldi and O. Oinazza, An accurate fractional Brownian
motion generator, \href{https://doi.org/10.1016/0378-4371(94)90531-2}{Physica
A \textbf{208}, 21 (1994)}.

\bibitem{beck2003} C. Beck and E. G. D. Cohen, Superstatistics,
\href{https://doi.org/10.1016/S0378-4371(03)00019-0}{Physica (Amsterdam)
\textbf{322A}, 267 (2003)}.

\bibitem{weis2013}  M. Weiss, Single-particle tracking data
reveal anticorrelated fractional Brownian motion in crowded fluids,
\href{https://doi.org/10.1103/PhysRevE.88.010101}{Phys. Rev. E \textbf{88},
010101 (2013)}.

\bibitem{weig2011} A. V. Weigel, B. Simon, M. M. Tamkun, and D. Krapf,
Ergodic and Nonergodic Processes Coexist in the Plasma Membrane as Observed by
Single-Molecule Tracking, \href{https://doi.org/10.1073/pnas.1016325108}{Proc.
Natl. Acad. Sci. \textbf{108}, 6438 (2011)}.

\bibitem{siko2017} G. Sikora, K. Burnecki, and A. Wy{\l}oma{\'n}ska,
Mean-squared-displacement statistical test for fractional Brownian motion,
\href{https://doi.org/10.1103/PhysRevE.95.032110}{Phys. Rev. E \textbf{95},
032110 (2017)}.

\bibitem{vester} C. L. Vestergaard, P. C. Blainey, and H. Flyvbjerg, Optimal
estimation of diffusion coefficients from single-particle trajectories,
Phys. Rev. E \textbf{89}, 022726 (2014).

\bibitem{kevin} A. Robson, K. Burrage, and M. C. Leake Inferring diffusion
in single live cells at the single-molecule level, Phil. Trans. R. Soc. Lond.
B \textbf{368}, 20120029 (2013).

\bibitem{michael} J. Krog, L. H. Jacobsen, F. W. Lund, D. W{\"u}stner, and
M. A. Lomholt, Bayesian model selection with fractional Brownian motion,
J. Stat. Mech. \textbf{2018}, 093501 (2018).

\bibitem{thap2018} S. Thapa, M. A. Lomholt, J. Krog, A. G. Cherstvy,
R. Metzler, Bayesian analysis of single-particle tracking
data using the nested-sampling algorithm: maximum-likelihood
model selection applied to stochastic-diffusivity data,
\href{https://doi.org/10.1039/C8CP04043E}{Phys. Chem. Chem. Phys. \textbf{20},
29018 (2018)}.

\bibitem{gorka} Gorka Mu\~{n}oz-Gil, M. A. Garcia-March, C. Manzo, J. D.
Mart{\'i}n-Guerrero, and M. Lewenstein, Single trajectory characterization
via machine learning, New J. Phys. \textbf{22}, 013010 (2020).

\bibitem{seck2023}  H. Seckler, J. Szwabi{\'n}ski, and R. Metzler,
Machine-learning solutions for the analysis of single-particle diffusion
trajectories, \href{https://doi.org/10.1021/acs.jpclett.3c01351}{J. Phys.
Chem. Lett. \textbf{14}, 7910 (2023)}.

\bibitem{manzo2021} C. Manzo, Extreme learning machine for the
characterization of anomalous diffusion from single trajectories
(AnDi-ELM), \href{https://doi.org/10.1088/1751-8121/ac13dd}{J. Phys. A:
Math. Theor. \textbf{54}, 334002 (2021)}.

\bibitem{seck2022}  H. Seckler and R. Metzler, Bayesian deep
learning for error estimation in the analysis of anomalous diffusion,
\href{https://doi.org/10.1038/s41467-022-34305-6}{Nat. Commun. \textbf{13},
6717 (2022)} .

\bibitem{szwa2019} P. Kowalek, H. Loch-Olszewska,
and J. Szwabi{\'n}ski, Classification of diffusion modes in
single-particle tracking data: Feature-based versus deep-learning approach,
\href{https://doi.org/10.1103/PhysRevE.100.032410}{Phys. Rev. E \textbf{100},
032410 (2019)}.

\bibitem{gork2021} G. Mu{\~n}oz-Gil et al., Objective
comparison of methods to decode anomalous diffusion,
\href{https://doi.org/10.1038/s41467-021-26320-w}{Nat. Commun. \textbf{12},
6253 (2021)}.

\bibitem{gran2019} N. Granik, L. E. Weiss, E. Nehme,
M. Levin, M. Chein, E. Perlson, Y. Roichman, and Y. Shechtman,
Single-particle diffusion characterization by deep learning.
\href{https://doi.org/10.1016/j.bpj.2019.06.015}{Biophys. J. \textbf{117},
185 (2019)}.

\bibitem{han2020} D. Han, N. Korabel, R. Chen, M. Johnston,
A. Gavrilova, V. J. Allan, S. Fedotov, and T. A. Waigh, Deciphering
anomalous heterogeneous intracellular transport with neural networks,
\href{https://doi.org/10.7554/eLife.52224}{eLife \textbf{9}, e52224 (2020)}.

\bibitem{balc25} M. Balcerek, A. Pacheco-Pozo, A. Wy{\l}oma{\'n}ska, and
D. Krapf, Evaluating Gaussianity of heterogeneous fractional Brownian motion,
\href{	 https://doi.org/10.48550/arXiv.2501.10472}{arXiv:2501.10472}.

\bibitem{yann22} Y. Lanoisel\'ee, A. Stanislavsky, D. Calebiro, and
A. Weron, Temperature and friction fluctuations inside a harmonic potential,
\href{https://doi.org/10.1103/PhysRevE.106.064127}{Phys. Rev. E \textbf{106},
064127 (2022)}.

\bibitem{vitt2024a} V. Sposini, S. Nampoothiri,
A. Chechkin, E. Orlandini, F. Seno, and F. Baldovin, Being
heterogeneous is disadvantageous: Brownian non-Gaussian searches,
\href{https://doi.org/10.1103/PhysRevE.109.034120}{Phys. Rev. E \textbf{109},
034120 (2024)}.

\bibitem{vitt2024b} V. Sposini, S. Nampoothiri, A. Chechkin,
E. Orlandini, F. Seno, and F. Baldovin, Being heterogeneous
is advantageous: extreme Brownian non-Gaussian searches,
\href{https://doi.org/10.1103/PhysRevLett.132.117101}{Phys. Rev. Lett.
\textbf{132}, 117101 (2024)}.

\bibitem{evan11} M. R. Evans and S. N. Majumdar, Diffusion with Stochastic
Resetting, \href{https://doi.org/10.1103/PhysRevLett.106.160601}{Phys. Rev.
Lett. \textbf{106}, 160601 (2011)}.

\bibitem{pal17} A. Pal and S. Reuveni, First Passage Under Restart,
\href{https://doi.org/10.1103/PhysRevLett.118.030603}{Phys. Rev.
Lett. \textbf{118}, 030603 (2017)}.

\bibitem{stella1} L. Stella, A. Chechkin, and G. Teza, Anomalous dynamical
scaling determines universal critical singularities, Phys. Rev. Lett.
\textbf{130}, 207104 (2023).

\bibitem{stella2} L. Stella, A. Chechkin, and G. Teza, Universal singularities
of anomalous diffusion in the Richardson class, Phys. Rev. E \textbf{107},
054118 (2023).

\end{thebibliography}
\end{document}